\documentclass{article}

\usepackage[utf8]{inputenc}
\usepackage[T1]{fontenc}

\PassOptionsToPackage{table, dvipsnames}{xcolor}
\usepackage{xcolor}

\PassOptionsToPackage{colorlinks=true, citecolor=blue, linkcolor=blue, urlcolor=black}{hyperref}

\usepackage{arxiv}
\usepackage[utf8]{inputenc}
\usepackage[T1]{fontenc}
\usepackage{url}
\usepackage{booktabs}
\usepackage{nicefrac}
\usepackage{microtype}
\usepackage{lipsum}
\usepackage{graphicx}
\usepackage{natbib}
\usepackage{doi}
\usepackage{algorithm}
\usepackage{algpseudocode}%
\usepackage{listings}%
\usepackage{multirow}
\usepackage{multicol}
\usepackage{amsmath,amssymb,amsfonts}
\usepackage{amsthm}
\usepackage{mathrsfs}
\usepackage{float}
\usepackage{bm}
\usepackage{caption}
\usepackage{hyperref}

\usepackage{tabularx}
\usepackage{array}
\usepackage{pdflscape}

\title{Shallow-to-deep velocity model building via diffusion models-Part I: Method and Proof of concept}

\author{
	\href{https://orcid.org/0000-0001-8868-7967}{\includegraphics[scale=0.06]{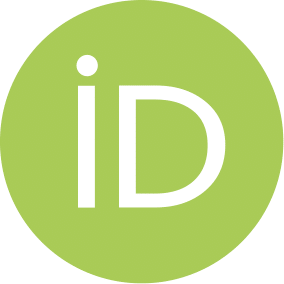}\hspace{1mm}Shijun~Cheng} \\
	Division of Physical Science and Engineering\\
	King Abdullah University of Science and Technology\\
	Thuwal 23955-6900, Saudi Arabia \\
        \And
    {Randy Harsuko} \\
	Division of Physical Science and Engineering\\
	King Abdullah University of Science and Technology\\
	Thuwal 23955-6900, Saudi Arabia \\
        \And
    {Tariq Alkhalifah} \\
	Division of Physical Science and Engineering\\
	King Abdullah University of Science and Technology\\
	Thuwal 23955-6900, Saudi Arabia \\
    [3ex]
  $^{*}$Corresponding author: \textbf{Shijun Cheng}~(\texttt{sjcheng.academic@gmail.com})
}

\renewcommand{\shorttitle}{Shallow-to-deep VMB: Part I}

\hypersetup{
pdftitle={A template for the arxiv style},
pdfsubject={q-bio.NC, q-bio.QM},
pdfauthor={David S.~Hippocampus, Elias D.~Striatum},
pdfkeywords={First keyword, Second keyword, More},
}

\begin{document}
\maketitle

\begin{abstract}
Seismic velocity model building (VMB) is fundamental for understanding subsurface structures. Traditional methods demand high-quality starting models and, also, remain limited in resolution in coverage and computationally intensive. Recent generative diffusion model-based approaches capture statistical priors to support traditional inversion methods, but these approaches do not account for the top to bottom progression of information (layer stripping) involved in surface recorded data, where deep velocity information depends on the shallow. To address this issue, we propose a depth-progressive diffusion framework that constructs velocity models incrementally from shallow to deep by propagating prior information. Our method trains on paired shallow-deep velocity patches with variable overlap and explicit depth encoding, integrating multiple geophysical constraints including well logs and seismic images (representing structural information). During inference, we synthesize overlapping depth slices using a progressive algorithm and merge them with Gaussian-weighted blending to eliminate boundary artifacts. This approach leverages both learned geological distributions and observed shallow priors while providing uncertainty quantification. Extensive numerical experiments on in-distribution tests and an out-of-distribution test demonstrate excellent VMB accuracy with a strong correlation between predicted uncertainty and actual errors. As a proof of concept, this part I employs idealized structural constraints derived from vertical reflectivity to validate the methodological framework. The companion paper (Part II) extends the approach to realistic structural constraints relying on migrated images with field data applications.
\end{abstract}

\section{\textbf{Introduction}}
Accurate seismic velocity models are essential tools for revealing Earth's internal structure, providing critical constraints for understanding crustal architecture, lithospheric evolution processes, and geodynamic mechanisms \citep{dziewonski1981preliminary, fichtner2010full, ritsema2011s40rts}. In the field of earthquake monitoring and early warning, high-precision velocity models directly affect the accuracy of earthquake location and the reliability of magnitude estimation \citep{wang2020regularized}. Furthermore, seismic velocity models play a vital role in resource exploration applications, particularly in seismic imaging for hydrocarbon exploration \citep{virieux2009overview}. Therefore, constructing high-precision, high-resolution seismic velocity models represents a fundamental and crucial requirement for the seismological community.

Traditional velocity model building (VMB) typically follows a two-step workflow, velocity analysis followed by tomography/migration analysis \citep{yilmaz2001seismic}. First, in common midpoint (CMP) gathers, reflection and refraction arrivals are picked either manually or automatically. The stacking velocity curves are extracted using semblance or spectral-analysis techniques and then converted into interval velocities using the Dix formula \citep{dix1955seismic}. These velocities are gridded in the horizontal and vertical directions to form an initial layered model \citep{alkhalifah1995velocity}. Subsequently, the initial model is iteratively updated through travel time tomographic inversion or residual moveout analysis on residual gathers, correcting the velocity model by minimizing travel time residuals or migration residuals \citep{luo1991wave, pratt1991combining, alkhalifah2003tau, sava2004wave, min2006refraction, shen2008automatic, van2010correlation, ma2013wave}. 

Although traditional tomography and residual moveout velocity analysis methods demonstrate robust performance in constructing initial velocity models, they often utilize only travel time information and struggle to resolve high-frequency, small-scale structures \citep{mora1989inversion}. Pure travel time inversion cannot fully exploit seismic amplitude and phase information, limiting its resolution and accuracy for imaging fine-scale geological features. To overcome these limitations, full-waveform inversion (FWI) minimizes amplitude and phase misfit between observed and synthetic waveforms, recovering much finer velocity variations \citep{tarantola1984inversion, tarantola1986strategy, choi2012application}. In typical implementations, FWI employs a multiscale strategy, where low frequencies are inverted first and progressively higher frequencies are added, to mitigate cycle-skipping risks \citep{bunks1995multiscale}, while utilizing Tikhonov smoothing or total variation regularization to suppress excessive oscillations in high-wavenumber components \citep{guitton2012blocky, peters2017constraints, esser2018total, kalita2019regularized}. Parallel advances in wave-equation tomography embed the full wave equation into the objective function, iteratively updating the model to reduce waveform residuals and achieving notable gains in both resolution and stability \citep{boonyasiriwat2009efficient}. However, both approaches demand a high-quality starting model and sufficient low-frequency content to avoid cycle-skipping, and they remain computationally intensive when repeatedly simulating wavefields over large domains \citep{virieux2009overview}. These practical challenges have motivated the search for alternative strategies that can both stabilize the inversion and accelerate VMB.

To address these limitations, data-driven methods based on deep learning have emerged as promising complements for traditional VMB workflows \citep{yu2021deep, mousavi2022deep}. Neural networks (NNs), particularly U‑Net architectures, can be trained end-to-end to map raw seismic gathers directly into the corresponding velocity model \citep{yang2019deep, wu2019inversionnet, li2019deep, liu2021deep, zhang2021deep, kazei2021mapping, du2022deep}. Once the network is trained, it can invert raw seismic data and directly produce a fully resolved velocity model, entirely bypassing the need for an explicit starting model. This capability not only streamlines the workflow but also significantly reduces computational overhead. Meanwhile, NNs have been employed to extrapolate missing low-frequency content, providing synthetic low-frequency bands for FWI to mitigate cycle-skipping before introducing high frequencies \citep{ovcharenko2019deep, sun2020extrapolated, fang2020data, cheng2024self}. Complementing this are learned objective functions obtained through training an NN, which replace classical $l_2$ waveform errors to better capture phase consistency and subtle amplitude differences, accelerating convergence and enhancing robustness to noise and cycle-skipping \citep{sun2020ml, yang2023fwigan, saad2024siamesefwi}. Along another pathway, physics-informed neural networks (PINNs) and neural operators were originally developed to address the computational inefficiency of traditional forward solvers by learning surrogate wave propagators \citep{song2021wavefield, yang2021seismic, rasht2022physics, zhang2023seismic, yang2023rapid, zou2024deep}. Implicit FWI techniques go further, representing velocity models entirely as coordinate-based neural network weights, thereby eliminating dependence on initial background models and naturally achieving smoothing and multiscale representation through network architecture \citep{sun2023implicit, zhang2023multilayer, yang2025gabor}.

While these deep learning-based methods have achieved success, generative models have opened another avenue for VMB: learning rich prior distributions of the given velocity models. Generative adversarial networks (GANs) \citep{goodfellow2014generative} have been used to capture statistical distributions of geological scenarios from synthetic or measured data, enabling rapid sampling of geologically-constrained velocity models \citep{mosser2018rapid, zhang2019velocitygan, mosser2020stochastic, sun2023prestack}. Recently, diffusion model-based generators \citep{ho2020denoising} have demonstrated capabilities for refining or synthesizing entire velocity volumes, effectively painting high-resolution details consistent with learned patterns. These generative priors can play three major roles in the VMB workflow: (1) providing high-quality initial models for subsequent multiscale FWI or wave-equation tomography \citep{zhang2025diffvmb}; (2) serving as regularization terms in inversion objective functions, preventing overfitting and promoting geological plausibility \citep{wangfu2023, haysim2024}; (3) directly synthesizing velocity samples to generate training datasets required for downstream machine learning tasks \citep{wangfu2024}. 

Despite the successes of GANs and diffusion-based generators in capturing rich spatial priors for seismic velocity volumes, they exhibit several inherent limitations when applied to VMB. Adversarial training is prone to instability and mode collapse, which can leave portions of the velocity distribution undersampled. Diffusion models, while more robust, impose substantial memory and computational overhead during training and inference, since generating full‑scale velocity cubes in a single pass is rarely tractable. Critically, both frameworks rely solely on learned distribution sampling and, thus, do not enforce the natural top‑down progression of seismic information for surface seismic recorded data: both learn to synthesize full velocity cubes in one shot, without an explicit mechanism to propagate reliable shallow estimates into deeper layers. This deficiency prevents them from leveraging the depth‑dependent confidence gradient, where shallow velocities are well constrained by data but deeper regions remain underdetermined. As a result, it limits their effectiveness in reconstructing deep structures.

In response, \cite{harsuko2025propagating} proposed VelocityGPT, an autoregressive Transformer decoder inspired directly by the GPT family’s language modeling paradigm. In natural language processing, GPT learns to predict the next word from all preceding words. Analogously, VelocityGPT treats each depth slice of a velocity volume as a token, conditioning the generation of deeper layers on the entirety of previously synthesized shallow layers. This top-down formulation is ideal for the vast majority of recorded seismic data, where recording takes place on the Earth's surface. However, because GPT architectures were originally optimized for one-dimensional linguistic sequences rather than two-dimensional spatial fields, they can struggle to capture the complex spatial correlations inherent in velocity models, especially for realistic, large scales where modeling long-range dependencies imposes heavy memory and computational burdens.

To overcome these limitations, we introduce a depth-progressive diffusion paradigm for seismic VMB that marries the autoregressive insight of VelocityGPT with the high-fidelity synthesis capabilities of generative diffusion models (GDMs). Rather than treating velocity model construction as a sequential token prediction task, we reformulate it as a conditional image generation problem where deeper velocity structures are progressively synthesized conditioned on reliable shallow priors. Our method introduces a unique training strategy for diffusion models that enables explicit propagation of shallow velocity information during the generation process. Specifically, during training, paired shallow-deep patches are sampled from full-scale models and encoded with depth positions and auxiliary constraints (e.g., well logs, structural maps), and the diffusion network is conditioned to denoise the deeper patch using the shallower patch as a condition. At inference, the full model is synthesized by iteratively generating overlapping depth slices and blending them with Gaussian weights to guarantee spatial continuity and eliminate stitching artifacts. This training and inference strategy enables the model to learn depth-wise information propagation and to leverage both the learned prior distribution of velocity fields and, also, actual shallow velocity measurements during synthesis. As a result, it can construct velocity models more accurately, thereby reducing uncertainty in VMB. As a proof of concept and since the quality of a structural description using a seismic image varies depending on the data, in our numerical experiments in this paper (Part I), we employ idealized structural constraints derived from vertical reflectivity computed directly from reference velocity models to validate the methodological framework and demonstrate its potential for VMB. We test this approach on in-distribution and out-of-distribution synthetic data. To bridge the gap toward practical applications, the companion paper (Part II) extends the framework to incorporate realistic structural constraints extracted from migrated seismic images and validates the method on field seismic data.

To summaries, this paper makes the following key contributions:
\begin{enumerate}
  \item \textbf{Depth-progressive diffusion framework.} We present a novel generative diffusion model that constructs seismic velocity volumes incrementally from shallow to deep by propagating prior information, in contrast to one-shot or patch-wise GDM approaches.
  \item \textbf{Conditional training strategy.} We design a training scheme that samples paired shallow-deep patches with randomized overlap, encodes depth positions, and integrates multiple geophysical constraints (e.g.\ well logs, structural maps), enabling the network to learn depth-wise information propagation.
  \item \textbf{Overlap-based inference with Gaussian blending.} We introduce a depth-progressive inference algorithm that synthesizes overlapping depth slices and merges them with Gaussian-weighted accumulation to ensure spatial continuity and eliminate boundary artifacts in large-scale models.
  \item \textbf{Fusion of learned priors and observed prior constraints.} Our method leverages both the learned prior distribution of velocity fields and actual shallow velocity priors during inference, yielding more accurate reconstructions and reducing uncertainty in deep regions.
  \item \textbf{Realistic model generation and uncertainty quantification.} Through extensive numerical experiments, we demonstrate that our approach can generate highly realistic velocity models that faithfully reproduce statistical and structural characteristics of subsurface data, while enabling efficient uncertainty estimation via ensemble sampling.
\end{enumerate}

\section{\textbf{Background}}
In this section, we first review the fundamentals of generative diffusion models (GDMs), including their forward noise-addition and learned reverse denoising processes, as well as common training objectives for noise versus data prediction. We then illustrate how GDMs have been applied to seismic velocity model synthesis and highlight the limitations of existing patch-wise generation methods, such as boundary artifacts, lack of depth-wise information propagation, and scalability challenges for large-scale volumes. These insights motivate our depth-progressive synthesis paradigm introduced in the following section.

\subsection{Generative diffusion models}
GDMs are a class of probabilistic deep generative models that learn to generate high-quality samples by modeling a gradual denoising process from pure noise to data. The framework consists of two essential components: a forward process that progressively adds noise to data, and a reverse process that learns to remove noise.

The forward process is a fixed Markov chain that gradually corrupts the original clean data $x_0$ by adding Gaussian noise over $T$ time steps, eventually transforming it into pure noise $x_T$:
\begin{equation}\label{eq1}
    q(x_{1:T}|x_0) = \prod_{t=1}^{T} q(x_t|x_{t-1}),
\end{equation}
where $q$ is a conditional distribution in which each transition step can be defined as:
\begin{equation}\label{eq2}
    q(x_t|x_{t-1}) = \mathcal{N}(x_t; \sqrt{1-\beta_t}x_{t-1}, \beta_t \mathbf{I}).
\end{equation}
Here, $\beta_t$ represents the predefined noise schedule, and $\mathbf{I}$ denotes isotropic Gaussian noise. Using the reparameterization trick, we can directly sample $x_t$ from $x_0$ at any timestep $t$:
\begin{equation}\label{eq3}
    x_t = \sqrt{\bar{\alpha}_t}x_0 + \sqrt{1-\bar{\alpha}_t}\epsilon,
\end{equation}
where $\bar{\alpha}_t = \prod_{s=1}^{t}(1-\beta_s)$, and $\epsilon \sim \mathcal{N}(0, \mathbf{I})$ represents Gaussian noise.

The reverse process aims to learn the denoising transitions that gradually recover the original data $x_0$ from noise $x_T$:
\begin{equation}\label{eq4}
    p_\theta(x_{0:T}) = p(x_T)\prod_{t=1}^{T} p_\theta(x_{t-1}|x_t),
\end{equation}
where each denoising step is modeled as:
\begin{equation}\label{eq5}
    p_\theta(x_{t-1} | x_t) = \mathcal{N}(x_{t-1}; \mu_\theta(x_t, t), \Sigma_\theta(x_t, t)).
\end{equation}
Here, $\mu_\theta(x_t, t)$ and $\Sigma_\theta(x_t, t)$ represent the mean and covariance of the denoising distribution, respectively, which need to be estimated through NNs. In practice, $\Sigma_\theta$ is typically set to a predetermined or fixed schedule, such as $\sigma_tz$, which can be used for stochastic sampling with $z \sim \mathcal{N}(0, \mathbf{I})$, while only $\mu_\theta(x_t, t)$ is learned through NN training.

A common approach is to train an NN to predict the added noise, where the mean is parameterized as:
\begin{equation}\label{eq6}
   \mu_\theta(x_t, t) = \frac{1}{\sqrt{\alpha_t}} \left( x_t - \frac{1 - \alpha_t}{\sqrt{1 - \bar{\alpha}_t}} \epsilon_\theta(x_t, t) \right).
\end{equation}
Alternatively, the NN can be trained to directly predict the clean data $x_0$, denoted as $x_{0,\theta}(x_t, t)$. In this case, the sampling process follows:
\begin{equation}\label{eq7}
    \mu_\theta(x_t, t) = \sqrt{\bar{\alpha}_{t-1}} x_{0,\theta}(x_t, t) + \sqrt{1 - \bar{\alpha}_{t-1}} \hat{\epsilon}(x_t, t),
\end{equation}
where $\hat{\epsilon}(x_t, t)$ is an estimate of the added noise using the predicted $x_{0,\theta}(x_t, t)$ as follows:
\begin{equation}\label{eq8}
   \hat{\epsilon}(x_t, t) = \frac{x_t - \sqrt{\bar{\alpha}_t} x_{0,\theta}(x_t, t)}{\sqrt{1 - \bar{\alpha}_t}}.
\end{equation}

The choice of network prediction target directly influences the training optimization objective. When the network is trained to predict the added noise $\epsilon$, the optimization objective minimizes the loss between the network prediction and the ground truth noise:
\begin{equation}\label{eq9}
   \mathcal{L}_{\epsilon} = \mathbb{E}_{x_0, \epsilon, t} \left[ \| \epsilon - \epsilon_\theta(x_t, t) \|^2 \right].
\end{equation}
Conversely, when the network directly predicts the clean data $x_0$, the optimization objective becomes minimizing the loss between the predicted and ground truth clean data:
\begin{equation}\label{eq10}
   \mathcal{L}_{x_0} = \mathbb{E}_{x_0, \epsilon, t} \left[ \| x_0 - x_{0, \theta}(x_t, t) \|^2 \right],
\end{equation}
Both training strategies enable the model to learn the underlying data distribution, though they differ in their explicit learning targets and may exhibit different convergence properties and generation quality.

\subsection{Conventional diffusion-based velocity model synthesis}
In conventional GDM-based seismic velocity model synthesis, velocity models are generated by adding controlled noise to realistic velocity model samples and conditioning the synthesis process on various geological and geophysical constraints \citep{wangfu2024, zhang2025diffvmb}. Typically, this involves embedding conditions to guide the velocity model generation process. The primary conditioning information in velocity model synthesis includes:
\begin{itemize}
   \item \textbf{Well logs}: Provide high-resolution local velocity constraints at specific spatial locations
   \item \textbf{Subsurface structure}: Can be reflectivity, seismic convolution images, or migrated images that provide geometric information about subsurface structures
   \item \textbf{Geological class}: Provides categorical geological information that guides the geological characteristics of generated models.
\end{itemize}
These conditions can be concatenated directly with the noisy input at the network's initial layer or integrated into each residual block within the network through dedicated embedding layers.

During training, the model takes a real velocity model $x_0$ and applies the forward diffusion process to obtain the noisy version $x_t$ as input to the network, where the network learns to predict the added noise or the real velocity model. Actually, the models can be trained conditionally, unconditionally, or through a mixed strategy. In unconditional training, the conditioning inputs are either set to zero or explicitly marked as None, allowing the network to learn solely from the intrinsic statistical distribution of the data without external constraints. Mixed training combines conditional and unconditional approaches, where the conditioning inputs are randomly omitted (set to zero or None) during some training iterations, helping the model become more robust and versatile. 

Once trained, the network can synthesize numerous velocity models during inference. The synthesis can be either conditional or unconditional, depending on whether we provide conditioning information such as well data and structural constraints. In conditional generation, specific conditions are provided to guide the generation, resulting in outputs that closely follow these constraints. Conversely, in unconditional generation, no conditions are supplied, allowing the network to sample purely from the learned data distribution without external guidance.

While conventional GDM-based velocity model synthesis has shown promising results, it faces several fundamental limitations when applied to realistic seismic velocity model building scenarios. First, conventional approaches typically generate the entire velocity model directly in a single pass. However, realistic seismic velocity models often have substantial dimensions especially in 3D, making direct training on full-scale models computationally prohibitive in terms of both memory consumption and training time. To address this computational challenge, a common alternative is to train a GDM on small velocity model patches, like $64\times64$. During inference, since the quality of diffusion models degrades significantly when generating models beyond their training dimensions, the common practice is to divide the full-scale velocity model synthesis into multiple patch-wise generations, each matching the training data dimensions, followed by subsequent merging. 

This patch-based approach, however, introduces several critical issues. When generating individual patches, the model lacks awareness of each patch's spatial context within the larger velocity model, making it difficult to capture realistic velocity distributions that span the entire domain. For instance, in realistic subsurface environments, velocities generally exhibit an increasing value trend with depth, a fundamental geological characteristic that patch-wise synthesis may fail to represent accurately. Moreover, the direct concatenation of independently generated patches often results in visible boundary artifacts and discontinuities in the final assembled velocity model.

Furthermore, conventional GDM-based synthesis relies solely on sampling from the learned velocity distribution with auxiliary conditional information during generation. In practice, however, we sometimes have access to reliable shallow velocity information, which can be as simple as the water later for marine data. Such information can constrain the velocity building process and reduce subsurface uncertainty \citep{harsuko2025propagating}. Current conventional approaches lack mechanisms to effectively incorporate such prior velocity information, missing an opportunity to leverage readily available constraints that could enhance the geological realism of the generated models. Even without shallow information, a framework that builds the model from shallow to deep accounts for the nature of information flow and confidence, and the fact that for surface recorded data the shallow part is better constrained by the data than the deep, so we often want to extend the learned shallow features to the deep. We will demonstrate this feature next.

\section{\textbf{Proposed method: Depth-progressive velocity model synthesis}}
We propose a novel paradigm for seismic velocity model building using GDMs that progressively builds velocity models from shallow to deep by propagating prior information. Unlike conventional approaches that generate entire velocity models or isolated patches, our method constructs velocity models incrementally in depth while simultaneously integrating multiple geophysical constraints. In the following, we will detail how our method propagates prior information from shallow to deep layers and seamlessly incorporates multiple geophysical constraints during both training and inference.

\subsection{Training strategy}
During training, we extract pairs of velocity model patches at different depths from full-scale velocity models. Each training sample consists of two patches with the following characteristics:
\begin{itemize}
    \item The lateral dimensions of these patches typically match the full lateral extent of the velocity model.
    \item The patches overlap in the depth direction with variable overlap size.
    \item The overlap size $o$ is randomly sampled during training:
    \begin{equation}\label{eq11}
        o \sim \mathcal{U}(1, n_z - 1),
    \end{equation}
where $n_z$ is the depth dimension of each patch and $\mathcal{U}$ denotes uniform distribution.
\end{itemize}

In this setup, the deeper velocity model patch $v_{deep} \in \mathbb{R}^{n_z \times n_x}$ serves as our prediction target $x_0$, while the shallower velocity model patch $v_{shallow} \in \mathbb{R}^{n_z \times n_x}$ acts as a conditioning input, where $n_x$ is the lateral dimension. Additional geophysical constraints, such as well velocities and structural information corresponding to the deeper patch, are also incorporated as conditions. During training, Gaussian noise is introduced into the deeper velocity patch through the diffusion process, resulting in a noisy version $x_t$ (see Equation (\ref{eq3})), which becomes the first input channel for the network. Meanwhile, the shallower velocity model patch is used as a prior conditional constraint as the second input channel for the network. Additionally, depth positions of grid points for both patches (denoted by $d_{shallow} \in \mathbb{R}^{n_z \times n_x}$ and $d_{deep} \in \mathbb{R}^{n_z \times n_x}$), matching the size of $x_t$ (our noisy $v_{deep}$), serve as the third and fourth input channels. Therefore, the network input consists of four channels 
\begin{equation}\label{eq12}
     \{x_t, v_{shallow}, d_{shallow}, d_{deep}\}. 
\end{equation}
Here, the depth encoding explicitly provides the absolute depth position for each grid point, enabling the network to learn depth-dependent velocity characteristics. 

In addition to the shallow velocity prior information constraints, we also use the same geophysical information constraints as in conventional GDM-based velocity model synthesis, such as well velocity and structure constraints. These constraints are encoded and then integrated into each residual layer of the network, which will be detailed in the subsequent network architecture section. Our training employs a mixed conditional and unconditional approach, randomly setting the well velocities $w$ and structural conditions $s$ to None with a probability of 5\%, as follows:
\begin{equation}\label{eq13}
    \mathbf{c} = 
    \begin{cases}
        \{w, s\} & \text{with probability } 0.95^2 \\
        \{\text{None}, s\} & \text{with probability } 0.05 \times 0.95 \\
        \{w, \text{None}\} & \text{with probability } 0.95 \times 0.05 \\
        \{\text{None}, \text{None}\} & \text{with probability } 0.05^2
    \end{cases}
\end{equation}
This strategy allows the model to handle scenarios with incomplete conditioning information during inference.

\begin{algorithm}[t]
\caption{Depth-progressive inference with Gaussian fusion}
\label{alg:depth_progressive}
\begin{algorithmic}[1]
\Require Trained diffusion model $p_\theta$ 
\Require Optional initial shallow velocity prior $v_{\text{init}}$ and geophysical constraints $(w,s)$
\Require Total depth $D$ and width $n_x$, patch depth $n_z$, stride $s=n_z/2$, Gaussian kernel parameter $\sigma$
\State Precompute $g(z)=\exp\!\big(-(z-n_z/2)^2/(2\sigma^2)\big)$ for $z=0,\ldots,n_z-1$
\State Initialize $V_{\text{result}}\leftarrow 0$, $V_{\text{weight}}\leftarrow 0 \in \mathbb{R}^{D\times n_x}$
\State Set $v_{\text{shallow}} \leftarrow v_{\text{init}}$ and place it at $[0:n_z)$:
\Statex ~~~~~~~~~~~~~~~~~~~~$V_{\text{result}}[0:n_z,:] {+}{=} g \odot v_{\text{shallow}}$;\quad $V_{\text{weight}}[0:n_z,:] {+}{=} g$
\State Build index set $\mathcal{I} = \{s,2s,\ldots,ks\} \cup \{D-n_z\}$
\For{each $i \in \mathcal{I}$}
    \State Build depth encodings $d_{\text{shallow}}, d_{\text{deep}}$
    \State Form input condition tensor $\{v_{\text{shallow}}, d_{\text{shallow}}, d_{\text{deep}}\}$
    \State Extract $(w_i, s_i)$ if available; else set to None
    \State Sample $x_T \sim \mathcal{N}(0,\mathbf{I})$
    \State Generate $v_i$ by reverse diffusion:
    \Statex ~~~~~~~~~~~~~~~~~~~~$v_i \leftarrow p_\theta(x_{0:T}) = p(x_T)\prod_{t=1}^T p_\theta(x_{t-1}|x_t, v_{shallow}, d_{shallow}, d_{deep}, w_i, s_i)$
    \State Update accumulators:
    \Statex ~~~~~~~~~~~~~~~~~~~~$V_{\text{result}}[i:i+n_z,:] {+}{=} g \odot v_i$
    \Statex ~~~~~~~~~~~~~~~~~~~~$V_{\text{weight}}[i:i+n_z,:] {+}{=} g$
    \State $v_{\text{shallow}} \leftarrow v_i$
\EndFor
\State \textbf{Return} $V = V_{\text{result}} \oslash V_{\text{weight}}$ \Comment{$\oslash$: element-wise division}
\end{algorithmic}
\end{algorithm}

\subsection{Depth-progressive inference}
After training, we generate the full velocity model through a depth-progressive synthesis process, as detailed in Algorithm~\ref{alg:depth_progressive}. The velocity model generation begins from shallow velocity priors $v_{init}$. In each inference step, the network predicts the velocity model at a deeper level, which maintains a predefined overlap with the shallower velocity model, typically half the depth of the velocity patch $n_z/2$. This newly predicted deeper velocity patch then serves as a conditioning shallow velocity input for the subsequent generation step. To ensure smooth transitions between overlapping regions and eliminate boundary artifacts, we employ a Gaussian-weighted accumulation strategy rather than simple concatenation. The iterative inference process continues until the complete velocity model depth is synthesized. The inference can be performed conditionally or unconditionally, depending on the availability of additional geophysical constraints such as well and structural information. 

For example, given a target velocity model of total depth $D$, we partition the generation task into multiple overlapping patches along the depth direction. The generation depth indices are defined as:
\begin{equation}\label{eq14}
    \mathcal{I} = \{s, 2s, 3s, ..., ks\} \cup \{D-n_z\},
\end{equation}
where $s = n_z/2$ is the step size ensuring overlap, $k = (D-n_z)/s$, and the final term $D-n_z$ ensures complete depth coverage when the model depth is not perfectly divisible by the step size. 

For each depth index $i \in \mathcal{I}$, we perform the following generation process:
\begin{enumerate}
    \item First, we prepare the depth encodings that explicitly indicate the absolute depth positions:
    \begin{align}
         d_{shallow} &= ([i-s, i-s+1, ..., i-s+n_z-1]^\text{T}).\text{repeat}(1, n_x), \\
         d_{deep} &= d_{shallow} + s,
    \end{align}
    where $[\cdot]^\text{T}$ denotes the column vector of depth indices, and $.\text{repeat}(1, n_x)$ replicates the depth values across the lateral dimension, resulting in a size $n_z \times n_x$. This ensures that each grid point contains its corresponding absolute depth position.
    \item Second, the shallow velocity condition $v_{shallow}$ (from the previous prediction or initial prior) is concatenated with these depth encodings:
    \begin{equation}
        \{v_{shallow}, d_{shallow}, d_{deep}\} \in \mathbb{R}^{3 \times n_z \times n_x}.
    \end{equation}
    \item Third, random noise $x_T$ is sampled from a standard Gaussian distribution:
    \begin{equation}
        x_T \sim \mathcal{N}(0, \mathbf{I}), \quad x_T \in \mathbb{R}^{1 \times n_z \times n_x}.
    \end{equation}
    \item Fourth, the sampled noise, together with the concatenated conditions, forms the input to our network. Additionally, if well or structural or both constraints are available, we extract the corresponding patches at depth $i$:
    \begin{align}
        w_i &= w[i:i+n_z, :] \quad \text{(well constraints)}, \\
        s_i &= s[i:i+n_z, :] \quad \text{(structural constraints)},
    \end{align}
    \item Finally, the trained diffusion model generates the velocity patch at depth index $i$:
    \begin{equation}
        v_i \leftarrow p_\theta(x_{0:T}) = p(x_T)\prod_{t=1}^{T} p_\theta(x_{t-1}|x_t, v_{shallow}, d_{shallow}, d_{deep}, w_i, s_i),
    \end{equation}
    where $p_\theta$ represents the trained diffusion model that iteratively denoises the input through the reverse diffusion process. This predicted patch $v_i$ serves as the shallow condition $v_{shallow}$ for the next depth iteration, enabling the progressive propagation of velocity information from shallow to deep.
\end{enumerate}

To achieve seamless blending between consecutive patches, we employ a Gaussian weighting scheme \citep{shi2024generative} to integrate the predicted velocity patch $v_i$ at each depth index $i \in \mathcal{I}$ into the full velocity model. The weight function is defined as:
\begin{equation}
    g(z) = \exp\left(-\frac{(z - n_z/2)^2}{2\sigma^2}\right), \quad z \in [0, n_z-1]
\end{equation}
where $\sigma$ controls the blending smoothness, which is commonly set as $n_z/8$ or $n_z/4$. This function assigns a maximum weight to the patch center and gradually decreases towards the boundaries. Throughout the generation process, we maintain two accumulation matrices:
\begin{align}
    V_{result} &\in \mathbb{R}^{D \times n_x} \quad \text{(accumulated weighted predictions)} \\
    V_{weight} &\in \mathbb{R}^{D \times n_x} \quad \text{(accumulated weights)}
\end{align}
For each generated patch $v_i$, we update the accumulators:
\begin{align}
    V_{result}[i:i+n_z, :] &\leftarrow V_{result}[i:i+n_z, :] + g \odot v_i \\
    V_{weight}[i:i+n_z, :] &\leftarrow V_{weight}[i:i+n_z, :] + g
\end{align}
where $\odot$ denotes element-wise multiplication and $w \in \mathbb{R}^{n_z \times 1}$ is broadcasted across the lateral dimension.

After completing all depth iterations, the final velocity model is obtained through element-wise normalization:
\begin{equation}
    V(z,x) = {V_{result}(z,x)}\oslash{V_{weight}(z,x)}
\end{equation}
where $\oslash$ represents element-wise division, and $z$ and $x$ denote the depth and lateral indices, respectively. This normalization ensures that each point in the final velocity model represents an optimal weighted combination of all contributing patches. For regions covered by multiple patches (i.e., overlapping zones), the Gaussian weighting naturally blends predictions, with greater influence from patch centers where the model has higher confidence. As a result, it effectively eliminates boundary artifacts while maintaining geological continuity throughout the depth profile.

\subsection{Uncertainty quantification}
Because the GDM is inherently probabilistic, different Gaussian noise realizations in the latent space yield different velocity realizations even under strong constraints (e.g., wells and structures), but the variance will be small, in this case, between the generated samples signaling high uncertainty. Therefore, we can estimate uncertainty by sampling multiple noise instances at each depth window and computing ensemble statistics.

Let $B$ be the number of independent noise samples drawn along the batch dimension for a given depth index $i$. For each $b\in\{1,\ldots,B\}$, we sample
\begin{equation}
x_T^{(b)} \sim \mathcal{N}(0,\mathbf{I}), \qquad x_T^{(b)} \in \mathbb{R}^{1\times n_z \times n_x},
\end{equation}
run the reverse diffusion, and obtain a velocity patch $v_i^{(b)}$. The element-wise ensemble mean and variance for this window are
\begin{equation}
\bar{v}_i = \frac{1}{B}\sum_{b=1}^{B} v_i^{(b)}, \qquad
\sigma_i^2 = \frac{1}{B}\sum_{b=1}^{B}\big(v_i^{(b)} - \bar{v}_i\big)^2 .
\end{equation}

After Gaussian blending across all depth windows (Algorithm~\ref{alg:depth_progressive}), we obtain $B$ full-size velocity volumes $\{V^{(b)}\}_{b=1}^B$. The final mean and variance fields over the entire model are
\begin{equation}
\bar{V}(z,x) = \frac{1}{B}\sum_{b=1}^{B} V^{(b)}(z,x), \qquad
\sigma_V^2(z,x) = \frac{1}{B}\sum_{b=1}^{B}\big(V^{(b)}(z,x) - \bar{V}(z,x)\big)^2 .
\end{equation}

In practice, we replicate the conditioning tensors (shallow patch, depth encodings, and optional geophysical constraints) along the batch dimension and feed $B$ different noise tensors through the same reverse process. The additional computational cost scales linearly with $B$.

\begin{figure*}[htbp]
\centering
\includegraphics[width=0.9\textwidth]{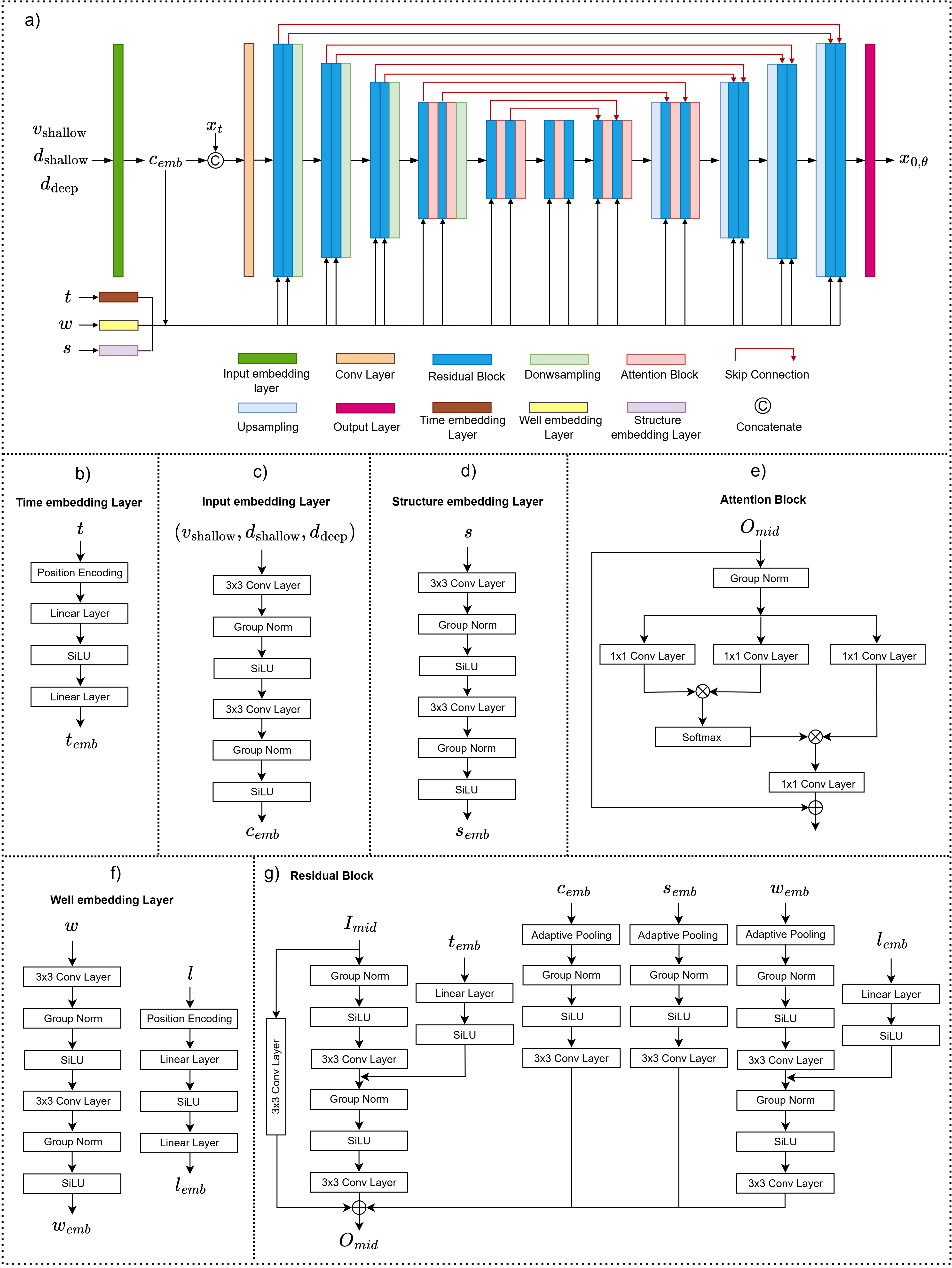}
\caption{An illustration of our network architecture. (a) The overall network structure. (b) Time embedding layer (where $t$ is the time step). (c) Input embedding layer (where $v_{shallow}$ is shallow velocity patch, and $d_{shallow}$ and $d_{deep}$ are depth positions of grid points for both patches, respectively). (d) Structure embedding layer (where $s$ is the structural image). (e) The attention block (where $O_{mid}$ is the block output in (g)). (f) Well embedding layer (where $w$ is the well velocity and $l$ is the well location). (g) The residual block. }
\label{fig1}
\end{figure*}

\subsection{Network architecture}
We employ a U-Net-based architecture specifically designed to handle multiple conditioning inputs and effectively integrate various geophysical constraints throughout the generation process. Figure \ref{fig1} illustrates our network architecture, with the overall baseline shown in panel (a) and key components detailed in panels (b-g).

For each training sample, the network accepts the following inputs: the noisy wavefield $x_t$, shallow velocity patch $v_{\text{shallow}}$, shallow depth position $d_{\text{shallow}}$, deep depth position $d_{\text{deep}}$, well data $w$, structural constraints $s$, and the diffusion timestep $t$. As described above, the three spatial priors $(v_{\text{shallow}}, d_{\text{shallow}}, d_{\text{deep}})$ are first concatenated and passed through an input-encoding module to produce a conditional feature embedding $c_{emb}$ (Figure \ref{fig1}c). The input‑encoding module applies a $3\times3$ convolution to project the concatenated spatial priors into 64 channels, followed by group normalization (GN) and activation via the sigmoid linear unit (SiLU). A second $3\times3$ convolution preserves the channel count, again followed by GN and SiLU, yielding $c_{emb}$. One branch of $c_{emb}$ is fused with the noisy input $x_t$ by channel‑wise concatenation and then passed through another $3\times3$ convolution to form the 64‑channel input tensor for the U‑Net. The other branch of $c_{emb}$ is injected into every residual block in both encoder and decoder, conditioning feature transforms on the spatial priors.

Our encoder-decoder comprises five resolution stages with feature widths $\{64,128,256,512,1024\}$. In the encoder’s first three stages, each applies two residual blocks (Figure \ref{fig1}g) followed by down‑sampling to capture increasingly abstract representations. From the fourth stage onward, we append a self‑attention layer after each residual block (Figure \ref{fig1}e) to enhance long‑range dependencies. The fifth stage omits further down‑sampling and instead feeds into a bottleneck sequence of (residual$\rightarrow$attention $\rightarrow$residual) blocks, maintaining 1024 channels. The decoder mirrors this layout in reverse, employing skip‑connections to re‑introduce features from the encoder at each corresponding stage.

To supply the network with explicit awareness of diffusion time, we embed the scalar timestep $t$ via a dedicated time‑encoding module (Figure \ref{fig1}b). We first apply a sinusoidal positional encoding to $t$, then project it through a linear layer, activate the output with SiLU, and project it again to produce a time embedding $t_{emb}$. This embedding is injected into every residual block, where it modulates feature normalization and thus conditions the denoising dynamics on the current diffusion step.

Meanwhile, in order to enable the network to effectively fuse geophysical information, we handle geological structure $s$ and well data $w$ with parallel encoding modules. The structure embedding layer (Figure \ref{fig1}d) mirrors the input‑encoding module to yield $s_{emb}$. For well data, we adopt a conventional sparse representation where well velocities are placed only at well locations while other positions are set to zero. This single-channel sparse well map is then processed through the same convolution$\rightarrow$GN$\rightarrow$SiLU pipeline to obtain a well-velocity embedding $w_{emb}$ (Figure \ref{fig1}f). Although this sparse representation provides both velocity and positional information, the predominance of zero entries can lead to weak constraints. To strengthen the well conditioning, we explicitly encode the well's spatial location $l$ through a time-like encoding pipeline (linear$\rightarrow$SiLU$\rightarrow$linear), producing a location embedding $l_{emb}$ that emphasizes where the well constraint should be applied. Both $w_{emb}$ and $l_{emb}$, together with $s_{emb}$, are injected into every residual block alongside $c_{emb}$ and $t_{emb}$.

To effectively fuse multiple conditioning inputs, we design an enhanced residual block (Figure \ref{fig1}g) that integrates five information streams: $t_{emb}$, $c_{emb}$, $s_{emb}$, $w_{emb}$, and $l_{emb}$. Specifically, the intermediate feature maps $I_{mid}$ are first normalized by GN, activated by SiLU, and convolved. In parallel, $t_{emb}$ is linearly projected and activated to produce a modulation vector. These outputs merge via scale-shift normalization, after which another GN$\rightarrow$SiLU$\rightarrow$conv sequence yields branch output $O_1$. A skip branch applies a single $3\times3$ convolution to $I_{mid}$, producing $O_2$. The conditional embeddings $c_{emb}$ and $s_{emb}$ are each adaptively average‑pooled to match the spatial dimensions of $I_{mid}$, then pass through GN$\rightarrow$SiLU$\rightarrow$conv to form $O_3$ and $O_4$. Likewise, $w_{emb}$ and $l_{emb}$ are pooled and fused using the same scale-shift scheme, producing $O_5$. Finally, the block output is the sum
\begin{equation}
    O_{mid} = O_1 + O_2 + O_3 + O_4 + O_5,
\end{equation}
which seamlessly integrates temporal, spatial, structural, and well constraints at every resolution. We emphasize that during both training and inference, each conditional branch may be disabled by setting its embedding to None. In such cases, that branch is bypassed and its output is treated as zero, ensuring that $O_{mid}$ remains effective even under unconditional or partially conditional scenarios. 

\section{\textbf{Numerical examples}}\label{sec:examples}

In the following, we present serval numerical experiments designed to evaluate the efficacy and robustness of our depth-progressive diffusion framework. First, we describe the industrial velocity-model datasets used for training and detail our training settings. Second, we assess performance on in-distribution tests, data drawn from the same distribution as the training set, including SEAM Arid, SEG/EAGE, and Overthrust models. Third, we demonstrate the model’s generalization capability through out-of-distribution (OOD) experiments on unseen velocity scenarios, i.e., the Marmousi model. 

\subsection{Training dataset and configuration}
 We train our diffusion model on a diverse collection of industrial velocity models, including both 2D sections and 3D volumes. The training dataset comprises ten benchmark models: seven 2D models (Otway, SEAM, BP1994, BP2004, BP2007, Hess, and Sigsbee) and three 3D volumes (SEAM Arid, SEG/EAGE, and Overthrust). For the 2D models, we first resize each model to 256 grid points in the depth direction. We then extract overlapping patches along the horizontal direction, with each patch having dimensions of $256 \times 256$ grid points. From each 2D model, we extract 83 training samples. For the 3D volumes, we extract multiple 2D slices from each volume. Each extracted slice is resized to 256 grid points in the vertical direction, and then overlapping patches of size $256 \times 256$ are sampled along the horizontal direction. From each 3D model, we extract 2430 training samples. To augment the training dataset, we apply horizontal flipping to all extracted patches, effectively doubling the dataset size. After data augmentation, our training dataset comprises 15742 velocity patches of size $256 \times 256$, providing diverse geological structures and velocity variations for robust model training.

During training, we extract paired shallow and deep patches of size $n_z=32$ (depth) by $n_x=256$ (lateral). From each deep patch we randomly select one vertical profile as a well-velocity constraint, where we consider the fact that well-log measurements are expensive and therefore sparsely available in practice. We, meanwhile, compute the vertical reflectivity of the deep patch to serve as a structural constraint. All inputs, including shallow and deep velocity patches, well velocities, and well locations, are normalized via min-max scaling to stabilize training.

Our diffusion network is trained with a forward process of $T=1000$ time steps and a cosine noise schedule. We employ an AdamW optimizer \citep{loshchilov2017fixing} with a fixed learning rate of $1\times10^{-4}$ and a batch size of 48. To further improve stability, we maintain an exponential moving average (EMA) of the network parameters with a rate of 0.999. The training runs for 400k iterations, which requires approximately 89 hours on a NVIDIA A100 GPU. At inference, each test velocity model of size $256 \times 256$ is synthesized in the depth-progressive manner described above. To accelerate inference, we adopt a denoising diffusion implicit model (DDIM) sampler \citep{song2020denoising} and reduce the number of reverse-diffusion steps to 10.

\begin{figure}[htbp]
\centering
\includegraphics[width=1\textwidth]{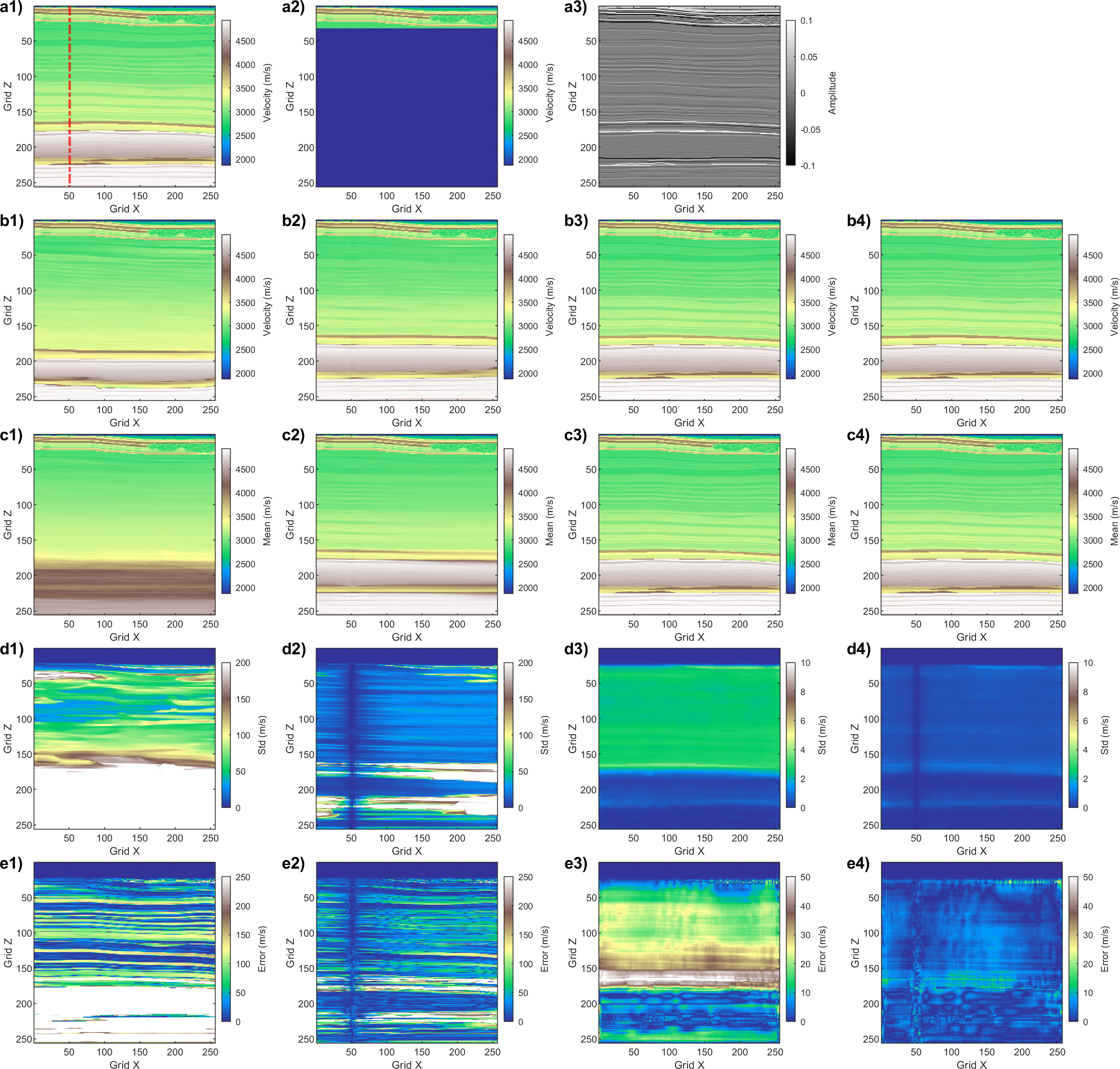}
\caption{Velocity model generation results for the SEAM Arid model. First row (a1-a3): true velocity model with well location (red dashed line), shallow prior, and vertical reflectivity used as structural constraint, respectively. Second row (b1-b4): individual realizations generated under unconditional, well-constrained, structure-constrained, and joint well and structure constrained, respectively. Third row (c1-c4): the mean of 50 realizations for the four scenarios. Fourth row (d1-d4): the standard deviation of 50 realizations. Fifth row (e1-e4): absolute difference between the mean (row c) and true model (a1).}
\label{fig2}
\end{figure}

\begin{figure}[htbp]
\centering
\includegraphics[width=0.6\textwidth]{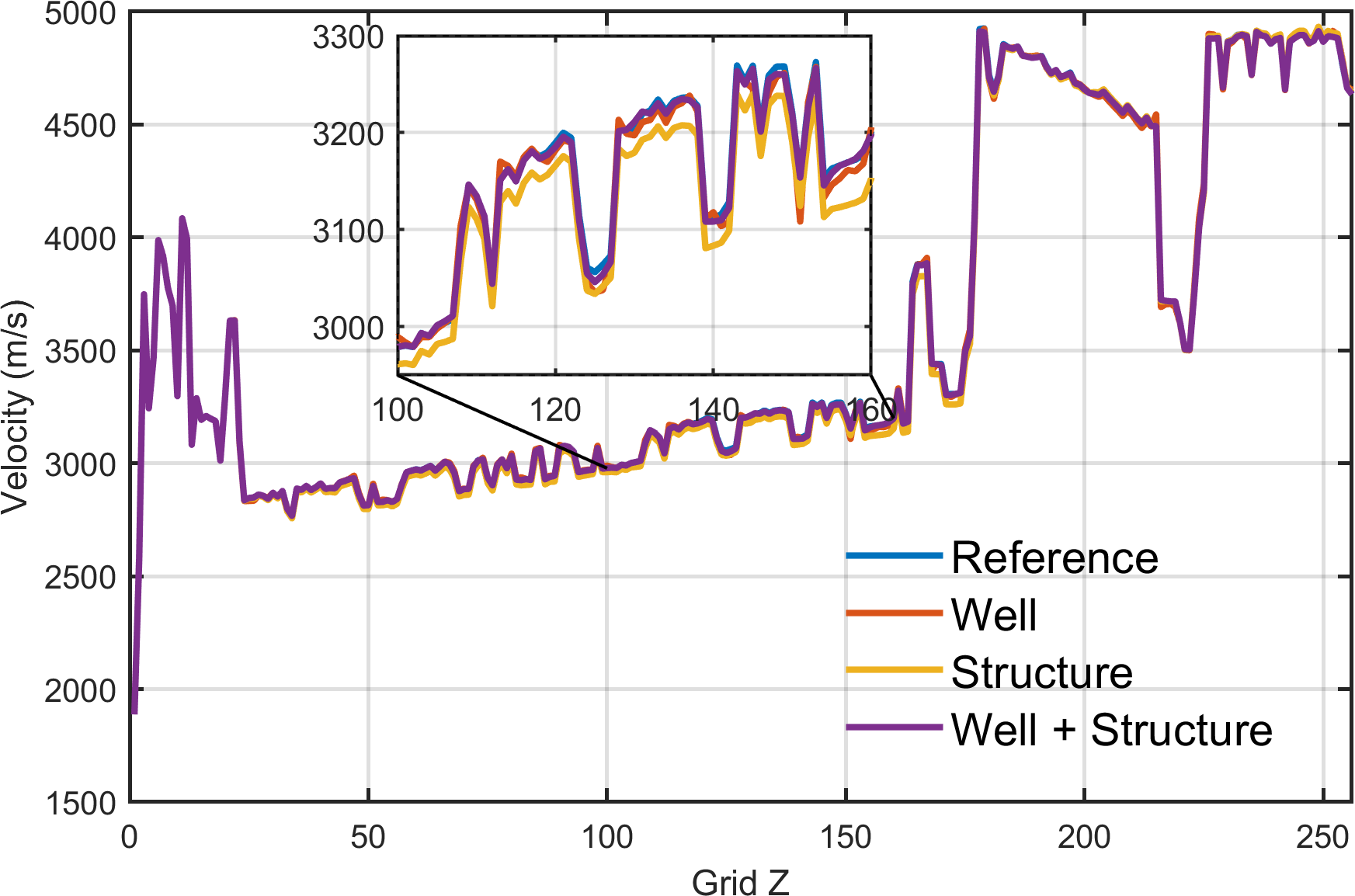}
\caption{Velocity profiles at the well location ($x=50$) for SEAM Arid: comparison of the reference velocity log (blue) with the mean predictions under well-only (orange), structure-only (yellow), and joint well and structure (purple) conditioning.}
\label{fig3}
\end{figure}

\subsection{In-distribution test: SEAM Arid}
We begin our evaluation by testing the proposed method on in-distribution models, where the test data shares similar geological characteristics and velocity distribution with the training dataset. This section presents the velocity model generation results on the SEAM Arid model, demonstrating the effectiveness of different constraint strategies in a controlled setting where the model has learned relevant prior distribution during training.

Figure \ref{fig2} illustrates the velocity model generation results for a test section extracted from the 3D SEAM Arid model. Panel (a1) displays the reference 2D velocity slice of size $256\times256$ grid points, where the red dashed line indicates the well location at $x=150$; panel (a2) shows the shallow prior velocity distribution extracted from the top 32 grid points; and panel (a3) presents the vertical reflectivity corresponding to the reference velocity model. The subsequent four rows present results under different constraint scenarios: a single realization (row b), the mean of 50 generated realizations (row c), the corresponding standard deviation (row d), and the absolute difference between the mean and reference model (row e). From left to right, each column corresponds to unconditional generation (column 1), well-constrained generation (column 2), structure-constrained generation (column 3), and jointly constrained (well and structure) generation (column 4).

The unconditional generation (Figure \ref{fig2}, column 1) demonstrates the capability of our model to generate geologically plausible velocity structures guided solely by the shallow prior. The single realization (b1) exhibits meaningful geological layering, while the mean model (c1) preserves general structural trends with a gradual velocity increase modulated by layered structures. This behavior stems from two key factors: first, the diverse training dataset enables the model to learn robust geological priors; second, the shallow velocity constraint (top $n_z=32$ grid points) provides essential initial guidance that propagates structural information into deeper layers through the depth-progressive synthesis process. Specifically, given only the shallow prior, the model identifies training samples with matching shallow characteristics (here primarily the SEAM Arid patches) and leverages the learned depth-dependent velocity distributions to progressively construct a plausible velocity model from top to bottom. This is a benefit of testing on an in-distribution sample, though not used in the training of the network. However, the standard deviation map (d1) reveals substantial uncertainty throughout the model domain, particularly in deeper regions. Correspondingly, the error map (e1) shows considerable differences from the reference model, indicating that unconditional generation alone is insufficient for accurate VMB without additional constraints.

Incorporating well constraints (Figure \ref{fig2}, column 2) provides localized velocity control and demonstrates significant improvements over unconditional generation. Both the single realization (b2) and mean model (c2) show substantially improved accuracy and more closely match the reference model. A particularly notable improvement is observed near the well location, where the velocity resolution is significantly enhanced with distinct velocity layers clearly visible throughout the depth range. The standard deviation map (d2) reveals a striking reduction in uncertainty, with near-zero values at the well location ($x=50$) forming a clear vertical zone of high confidence. Beyond this zone, uncertainty gradually increases but remains noticeably lower than the unconditional scenario. The error map (e2) demonstrates substantially reduced velocity differences, particularly pronounced in the deep layers and around the well. However, regions distant from the well still exhibit elevated errors, highlighting the limited lateral influence of a single sparse well.

The structure-constrained generation (Figure \ref{fig2}, column 3) demonstrates remarkable effectiveness in guiding VMB through vertical reflectivity. Both the single realization (b3) and mean model (c3) successfully capture the stratigraphic boundaries and lateral structural variations present in the reference model, with excellent agreement in velocity values across most of the domain. The standard deviation map (d3) exhibits a distinct pattern, with uncertainty concentrated along interfaces and structurally complex zones (e.g., depth $z=30\sim170$ with thin layers) rather than being uniformly distributed. The uncertainty magnitude is significantly reduced (display range: 0-10 m/s), indicating that structural constraints effectively reduce epistemic uncertainty. The error map (e3) confirms substantially reduced differences (display range: 0-50 m/s), with remaining errors primarily localized to thin-layered zones.

The jointly constrained generation (Figure \ref{fig2}, column 4) achieves the highest quality results by combining the complementary strengths of well and structural constraints. Both the single realization (b4) and mean model (c4) appear nearly identical to the reference model (a1), capturing both large-scale structural features and fine-scale velocity details. The standard deviation map (d4) reveals dramatic uncertainty reduction compared to using either constraint independently. The well constraint anchors absolute velocity values locally, while the structural constraint provides lateral continuity and guides transitions between layers. The error map (e4) demonstrates minimal differences across the entire model domain.

Another important observation, in particular, from Figure \ref{fig2} is the strong correlation between uncertainty maps (middle column) and difference maps (right column) across all scenarios. Regions with higher standard deviation almost correspond to areas with larger velocity errors relative to the reference model. This correlation validates the reliability of our uncertainty quantification framework and demonstrates that predicted uncertainties can serve as practical indicators of VMB confidence even when reference models are unavailable. This predictive capability represents a significant advantage of probabilistic generative modeling over deterministic approaches, enabling geophysicists to assess the reliability of synthesized velocity models and identify regions requiring additional data acquisition or constraints.

Figure \ref{fig3} further compares velocity profiles extracted at the well location ($x=50$) to illustrate how different constraint methods perform in reproducing the reference velocity. The velocity profiles clearly demonstrate the critical role of well constraints for local velocity calibration. Both the well-constrained model and the jointly constrained model achieve nearly perfect alignment with the reference velocity profile throughout the entire depth range, with velocity values precisely matching the well measurements. In contrast, the structure-constrained model exhibits slight deviations from the reference velocity, despite successfully capturing the overall structural trend and layer boundaries. This comparison underscores that well constraints are essential for calibrating absolute velocity values at well locations, while structural constraints excel at recovering geological interfaces and lateral continuity. The combination of both constraints leverages their complementary strengths: well data anchors accurate velocity values locally, and structural information propagates this accuracy laterally across the model domain.

\begin{figure}[htbp]
\centering
\includegraphics[width=1\textwidth]{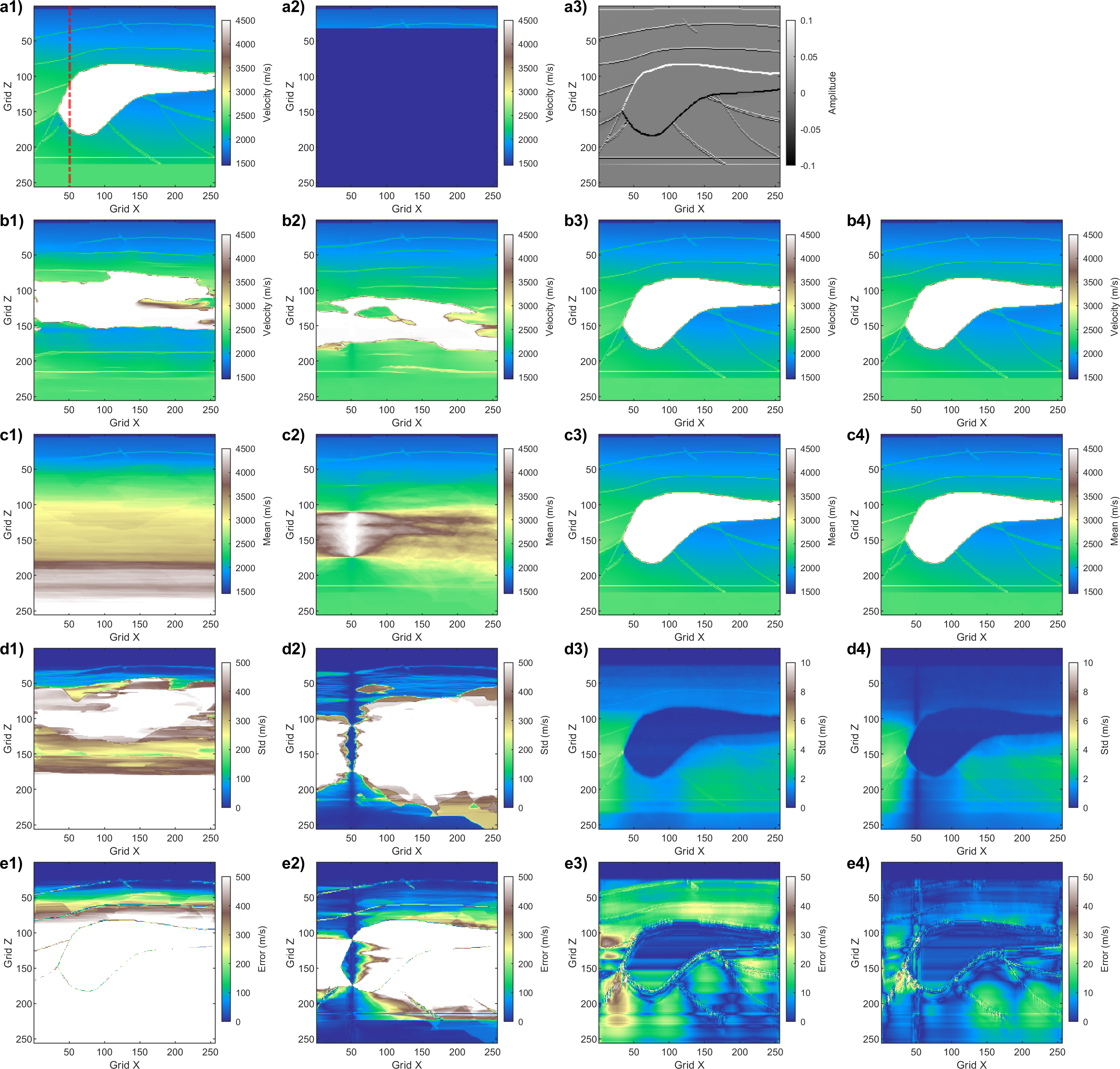}
\caption{Similar with Figure \ref{fig2}, but for SEG/EAGE model.}
\label{fig4}
\end{figure}

\begin{figure}[htbp]
\centering
\includegraphics[width=0.6\textwidth]{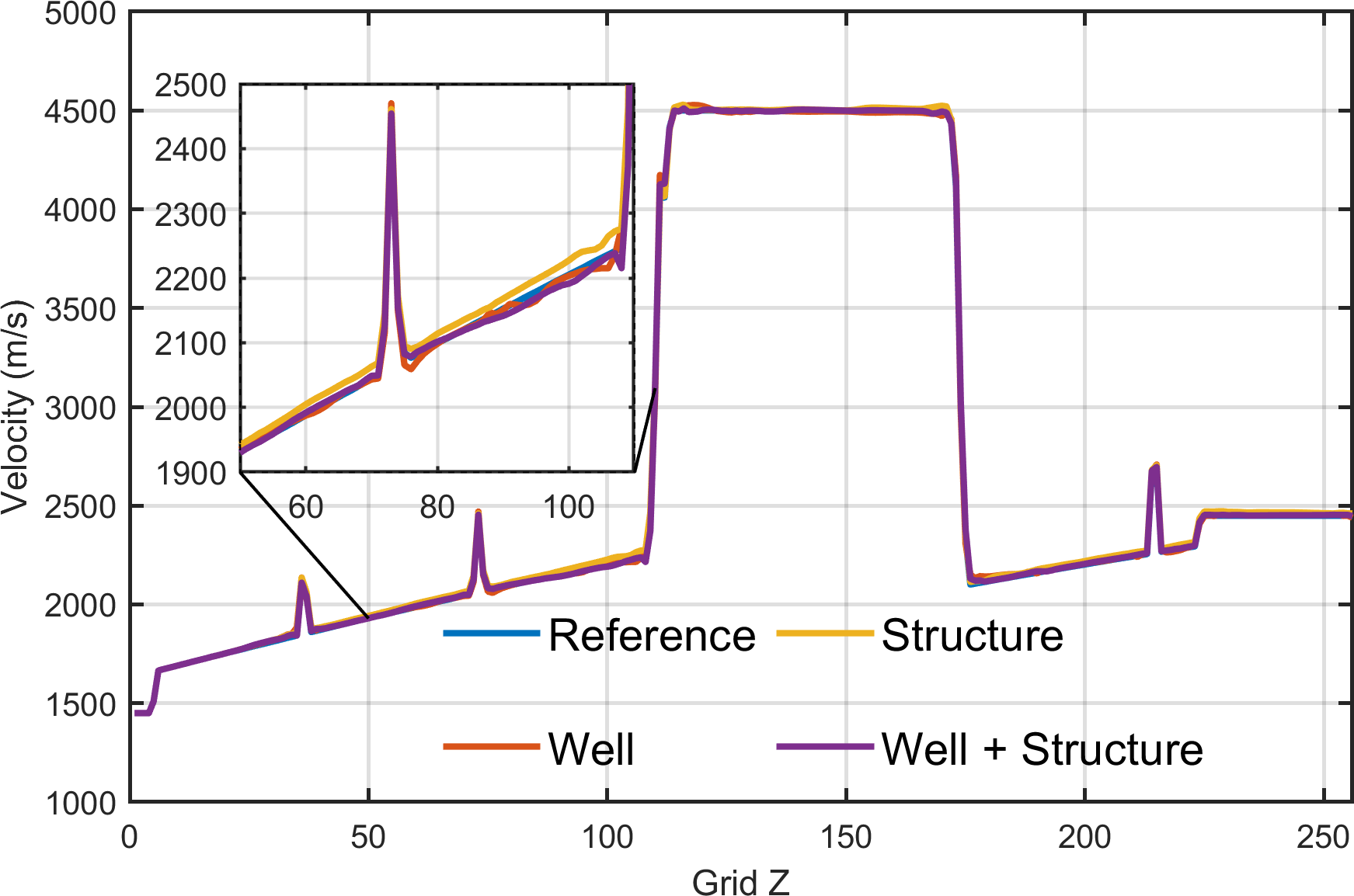}
\caption{Velocity profiles at the well location ($x=50$) for SEG/EAGE model: comparison of the reference velocity log (blue) with the mean predictions under well-only (orange), structure-only (yellow), and joint well and structure (purple) conditioning.}
\label{fig5}
\end{figure}

\subsection{In-distribution test: SEG/EAGE model}
We continue our in-distribution evaluation with a test section from the SEG/EAGE model, which presents a more challenging scenario due to the presence of a prominent salt body with sharp velocity contrasts (a more complex structure overall). Figure \ref{fig4} presents the velocity model generation results for a $256\times256$ test section from the SEG/EAGE model. The reference model (a1) features a salt body embedded within layered sediments, with the well location indicated at $x=50$. The figure follows the same organization as Figure \ref{fig2}: panels (a2) and (a3) show the shallow prior and vertical reflectivity, while rows b-e present single realizations, means, standard deviations, and errors for the four constraint scenarios (columns 1-4).

The unconditional generation (column 1) demonstrates an interesting capability: guided solely by the shallow prior, the model can approximately indicate the presence of a salt-like structure at roughly the correct depth range, as visible in both the single realization (b1). This suggests that the shallow velocity, considering our training set distribution, provides sufficient information to infer the approximate location of high-velocity anomalies through the learned geological priors. However, both the single realization (b1) and mean model (c1) exhibit substantial differences from the reference model. In fact, the mean model suggests that most of the generated samples might not include the salt body, or include it at none overlapping locations. However, the sample (b1) confirms that the salt body is part of the distribution and with the appropriate guidance, we should recover it at its expected values. The standard deviation map (d1) shows high uncertainty (up to 500 m/s), and the error map (e1) demonstrates large variance throughout the domain (display range: 0-500 m/s), but especially around the expected salt region and at depth. This result indicates that while unconditional generation can provide rough structural guidance, additional constraints are essential to improve VMB accuracy for complex geological features like salt bodies.

Incorporating well constraints (column 2) reduces VMB errors and uncertainties, particularly in areas near the well location. The models (b2, c2) show improved velocity accuracy around $x=50$, and the standard deviation map (d2) reveals substantially reduced uncertainty at the well. The error map (e2) demonstrates decreased velocity differences compared to the unconditional case, especially around the well. This result reaffirms the previous observation from the SEAM Arid test: well constraints are effective in calibrating velocities at the well location. However, regions distant from the well still exhibit elevated uncertainties and errors.

The structure-constrained generation (column 3) successfully reconstructs the salt body geometry with well-defined boundaries. Both the single realization (b3) and mean model (c3) closely match the reference model's salt structure and lateral extent. The standard deviation map (d3) shows significantly reduced uncertainty (display range: 0-10 m/s), concentrated primarily along salt boundaries. The error map (e3) reveals that errors within the salt body interior are relatively modest. However, noticeable errors (display range: 0-50 m/s) persist in the sedimentary layers surrounding the salt body. This observation suggests that while analytically computed structural constraints perform well at defining salt geometry, some velocity ambiguities remain in the transitional zones between different geological units.

The jointly constrained generation (column 4) achieves optimal results by combining both constraints. The models (b4, c4) accurately reconstruct salt geometry while maintaining precise velocity values throughout the domain. The standard deviation (d4) and error maps (e4) demonstrate dramatic reductions (display range: 0-10 m/s and 0-50 m/s respectively), with minimal uncertainties and errors across the entire model. The synergy between well and structural constraints effectively addresses the limitations observed in single-constraint scenarios. Similar to Figure \ref{fig2}, we can also see that strong spatial correlation exists between uncertainty and error maps across all scenarios, validating the reliability of our uncertainty quantification framework.

Figure \ref{fig5} compares velocity profiles at the well location ($x=50$). Both well-constrained and jointly constrained models achieve nearly perfect alignment with the reference throughout the entire depth range, accurately capturing the sharp velocity jump at the salt top and the high-velocity salt body. In contrast, the structure-constrained model exhibits noticeable velocity deviations. This comparison reinforces that structural constraints are essential for defining salt geometry, while well constraints are critical for calibrating absolute velocity values. Though the overall improvements to adding the well constraint to the structure seems small, the importance of the well will be better realized when the structural constraints are given by the RTM image, as we will see in part II. For now, the analytically computed structure maps implicitly contain information of velocity variation along interfaces.

\begin{figure}[htbp]
\centering
\includegraphics[width=1\textwidth]{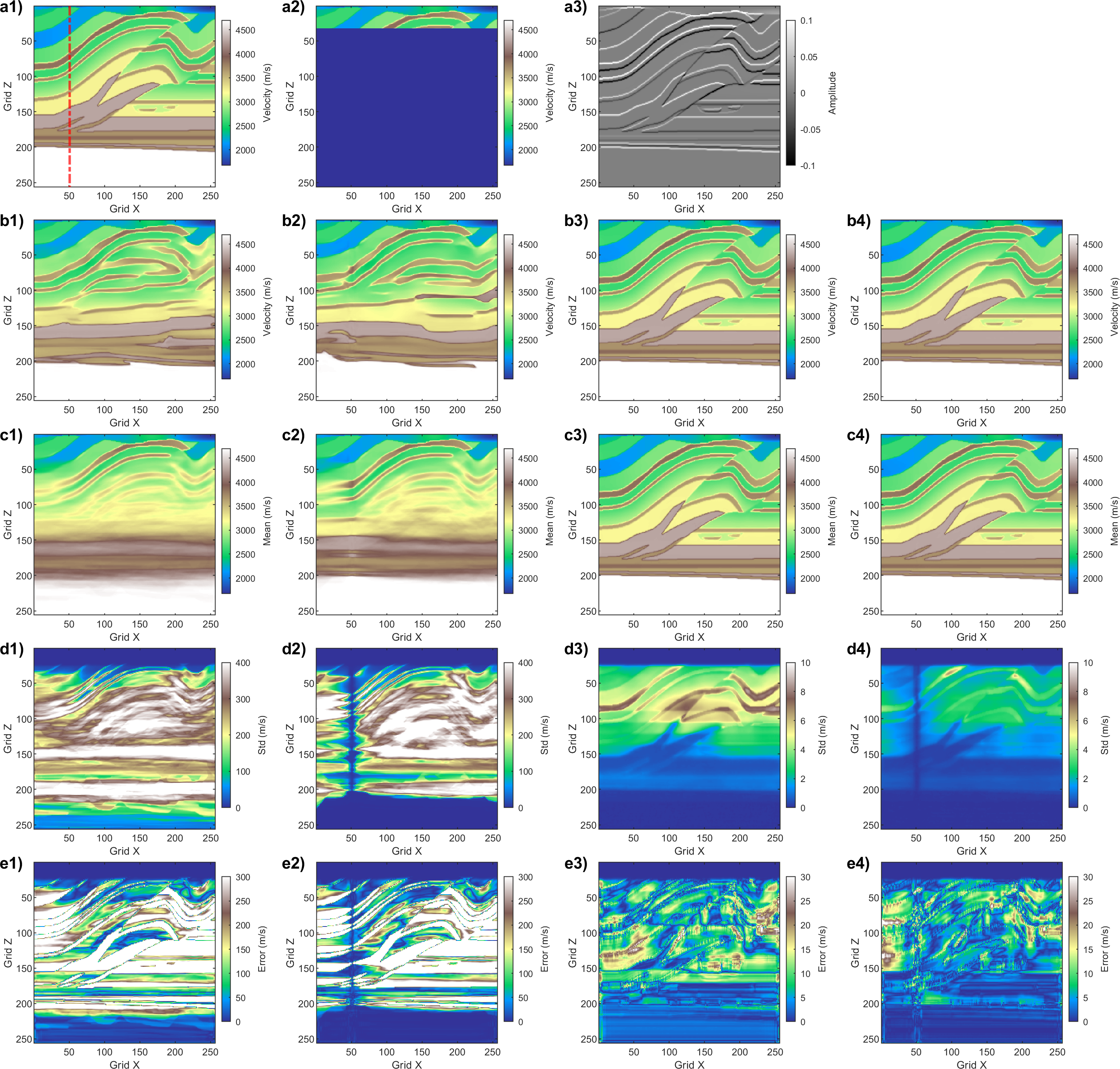}
\caption{Similar with Figure \ref{fig2}, but for Overthrust model.}
\label{fig6}
\end{figure}

\begin{figure}[htbp]
\centering
\includegraphics[width=0.6\textwidth]{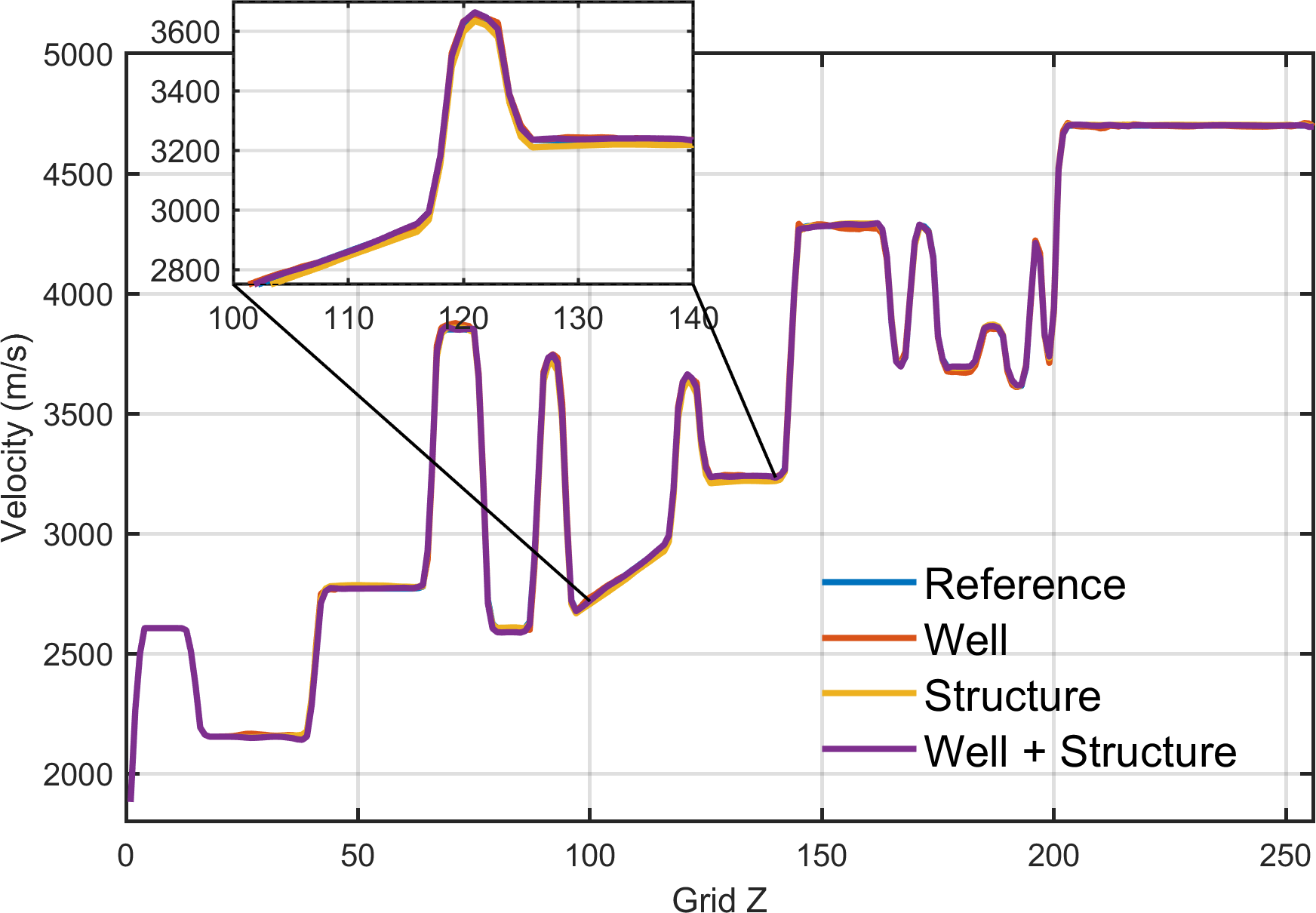}
\caption{Velocity profiles at the well location ($x=50$) for Overthrust model: comparison of the reference velocity log (blue) with the mean predictions under well-only (orange), structure-only (yellow), and joint well and structure (purple) conditioning.}
\label{fig7}
\end{figure}

\subsection{In-distribution test: Overthrust model}
Our final in-distribution test tackles a test section from the Overthrust model, which is characterized by two complex overthrust faults in the middle. Figure \ref{fig6} presents the velocity model generation results for a $256\times256$ test section from the Overthrust model. The reference model (a1) clearly shows the complex overthrust fault structures, with the well location indicated at $x=50$. The figure follows the same organization as Figure \ref{fig2}, with panels (a2) and (a3) showing the shallow prior and vertical reflectivity, and rows b-e presenting single realizations, means, standard deviations, and errors for the four constraint scenarios.

The unconditional generation (column 1) produces layered structures but fails to capture the complex fault geometry. Both the single realization (b1) and mean model (c1) exhibit overly smooth features compared to the reference model's structural complexity. The standard deviation (d1) and error maps (e1) reveal substantial uncertainties and differences throughout the domain, particularly in regions with steep dips and fault discontinuities.

Incorporating well constraints (column 2) improves velocity accuracy near the well location, as evidenced by reduced uncertainties (d2) and errors (e2) around $x=50$. However, the complex fault structures far from the well are not well constructed, reaffirming the limited lateral influence of sparse well data in structurally complex settings.

The structure-constrained generation (column 3) remarkably reconstructs the fault geometry throughout the domain. Both the single realization (b3) and mean model (c3) successfully capture the overthrust faults and structural discontinuities consistent with the reference model. The standard deviation map (d3) shows significantly reduced uncertainty (display range: 0-10 m/s), concentrated along fault zones. However, the error map (e3) reveals some velocity biases, indicating limitations in absolute velocity calibration despite excellent geometric reconstruction.

The jointly constrained generation (column 4) achieves optimal results by combining both constraints. The models (b4, c4) accurately represent both fault geometries and velocity distributions. The standard deviation (d4) and error maps (e4) demonstrate substantial reductions compared to single-constraint scenarios, with the well constraint effectively eliminating the velocity biases observed in the structure-only case.

Figure \ref{fig7} compares velocity profiles at the well location ($x=50$). Both well-constrained and jointly constrained models achieve nearly perfect alignment with the reference profile throughout the entire depth range, accurately capturing the complex velocity variations. The structure-constrained model shows noticeable deviations. This comparison once again confirms the complementary roles of structural and well constraints: structural information defines fault geometry and layer boundaries, while well data calibrates absolute velocity values at the well location.

\begin{figure}[htbp]
\centering
\includegraphics[width=1\textwidth]{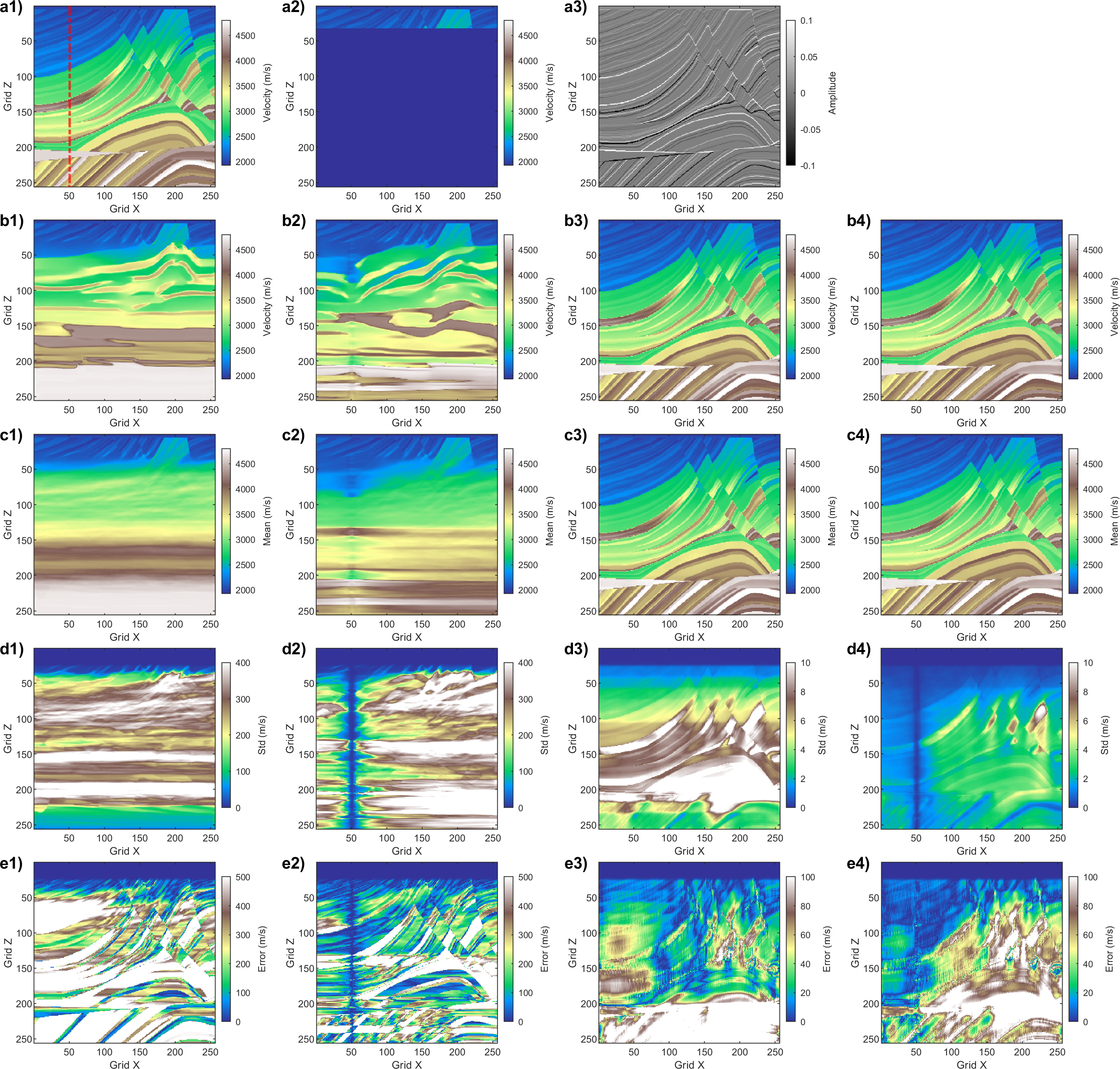}
\caption{Out-of-distribution velocity model generation results for the Marmousi model. The figure follows the layout of Figure \ref{fig2}.}
\label{fig8}
\end{figure}

\begin{figure}[htbp]
\centering
\includegraphics[width=0.6\textwidth]{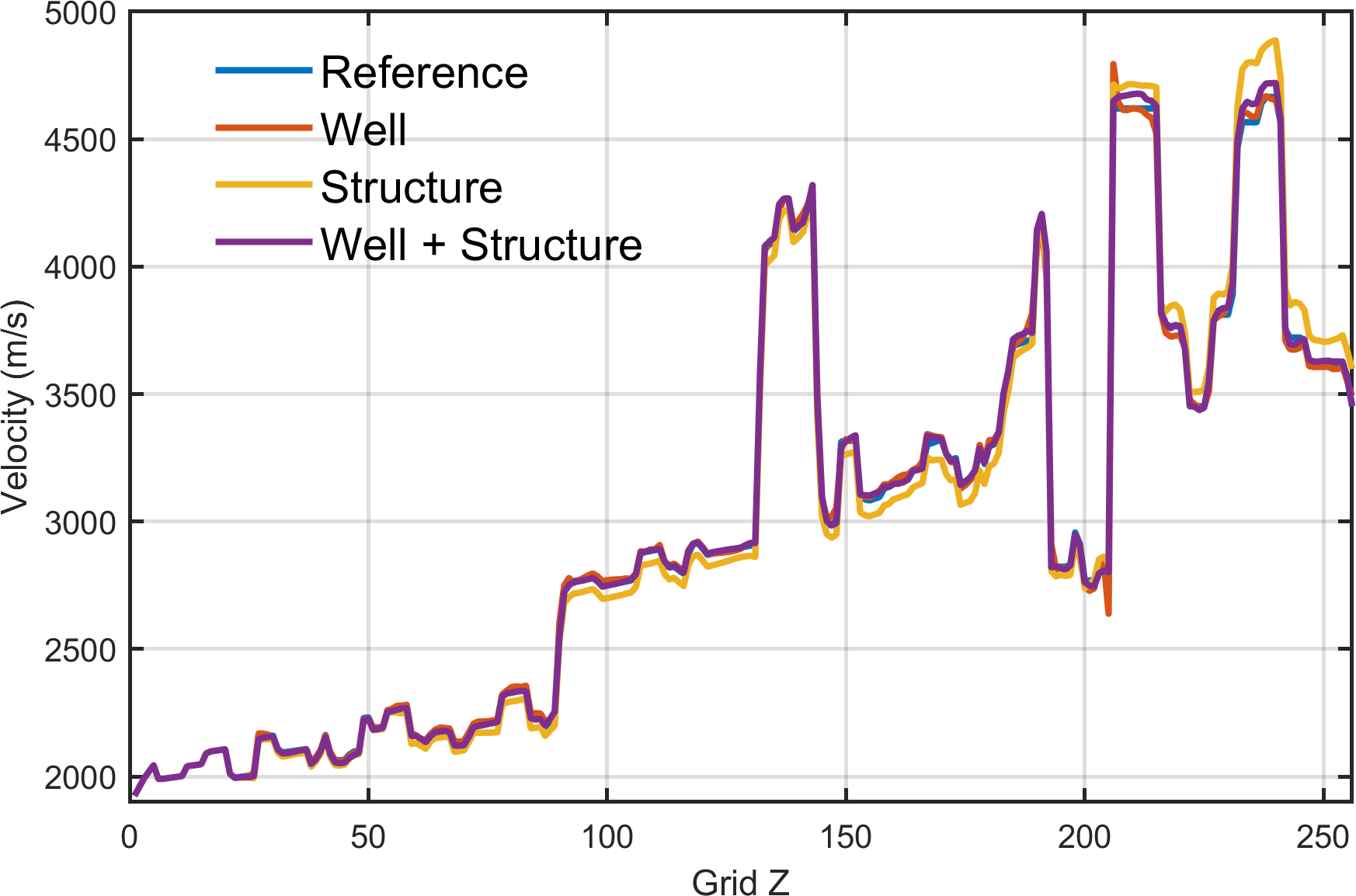}
\caption{Velocity profiles at the well location ($x=50$) for Marmousi model: comparison of the reference velocity log (blue) with the mean predictions under well-only (orange), structure-only (yellow), and joint well and structure (purple) conditioning.}
\label{fig9}
\end{figure}

\subsection{Out-of-distribution test: Marmousi model}
Finally, we evaluate our trained diffusion model on the Marmousi model, a challenging OOD scenario that reasonably differs from the training data in terms of velocity distribution and geological characteristics. Figure \ref{fig8} presents the velocity model generation results for a $256\times256$ test section from the Marmousi model. The reference model (a1) exhibits complex structures with a prominent fault and strong lateral velocity variations, with the well location indicated at $x=50$. The shallow prior and vertical reflectivity are shown in panels (a2) and (a3). The figure shares the same organization as previous figures, with rows b-e showing single realizations, means, standard deviations, and errors for the four constraint scenarios.

The unconditional generation (column 1) reveals an interesting behavior related to the shallow fault structure. The single realization (b1) exhibits features resembling the Overthrust model rather than the actual Marmousi reference. This occurs because the shallow prior (a2) contains fault information, and among the training samples, the Overthrust model patches contain the most abundant fault structures. Consequently, the model tends to generate velocity structures similar to Overthrust when guided solely by the shallow prior with fault characteristics. The mean model (c1) shows a more smoothed, layered structure resulting from averaging multiple realizations with Overthrust-like characteristics. The standard deviation map (d1) reveals substantial uncertainty throughout most of the domain, with relatively lower values only in the shallow section (where the prior provides guidance) and near the model bottom. The error map (e1) demonstrates large differences from the reference model, confirming that unconditional generation struggles with this OOD scenario.

Incorporating well constraints (column 2) effectively reduces both uncertainty and errors near the well location at $x=50$. The standard deviation map (d2) shows a clear vertical zone of reduced uncertainty centered at the well, and the error map (e2) demonstrates notable improvements around the well compared to the unconditional case. Interestingly, the single realization (b2) still exhibits Overthrust-like characteristics, suggesting that the well constraint primarily calibrates velocity values while the structural features remain influenced by the shallow fault information. However, the mean model (c2) shows improved resolution compared to the unconditional mean (c1), particularly at the well location where the well constraint introduces higher-resolution layering information.

The structure-constrained generation (column 3) successfully reconstructs the complex geometry of the Marmousi model. Both the single realization (b3) and mean model (c3) capture the major structural features and lateral variations present in the reference model. The standard deviation map (d3) shows significantly reduced uncertainty (display range: 0-10 m/s), and the error map (e3) demonstrates substantially improved accuracy (display range: 0-100 m/s). The performance is comparable to the in-distribution tests, indicating that structural constraints remain highly effective for geometry reconstruction even in OOD scenarios.

Again, the jointly constrained generation (column 4) achieves the best results among all constraint scenarios. The models (b4, c4) accurately reconstruct both the structural geometry and velocity distributions. Compared to the structure-only case, the joint constraints further reduce both uncertainty (d4) and errors (e4), consistent with the observations from the in-distribution tests. However, it must be acknowledged that compared to the in-distribution tests, the OOD results exhibit noticeably higher errors. This degradation in performance suggests that to achieve ideal accuracy across diverse geological settings, expanding the training distribution to include more varied velocity models would be beneficial.

Figure \ref{fig9} compares velocity profiles at the well location ($x=50$). Both well-constrained and jointly constrained models achieve excellent alignment with the reference profile throughout the entire depth range. In contrast, the structure-constrained model exhibits significantly larger deviations from the reference velocity profile compared to the in-distribution tests. These deviations are notably more pronounced than those observed in Figures \ref{fig3}, \ref{fig5}, and \ref{fig7}, where the structure-constrained profiles showed only slight differences from the reference. This suggests that the velocity variations along the interface for the Marmousi might have not been seen in the training set. This observation further confirms that in OOD scenarios, while structural constraints can ensure geometric accuracy, velocity calibration still requires well information to correct the velocity biases introduced by the distributional shift. The comparison underscores that well constraints become increasingly critical for achieving accurate velocity values when the test data diverges from the training distribution.

\section{\textbf{Understanding the depth-progressive framework}}
To provide deeper insights into our method, this section analyzes the fundamental mechanisms underlying depth-progressive velocity model building (VMB). We first introduce a data manifold perspective to explain how learned priors interact with observation constraints. We then demonstrate the roles of learned statistical distributions and shallow prior propagation using the Overthrust model example. Finally, we examine how shallow prior depth affects VMB performance. Together, these analyses reveal the underlying mechanisms that enable our method to generate geologically plausible velocity models, as well as the  limitations of our approach.

\begin{figure}[htbp]
\centering
\includegraphics[width=1\textwidth]{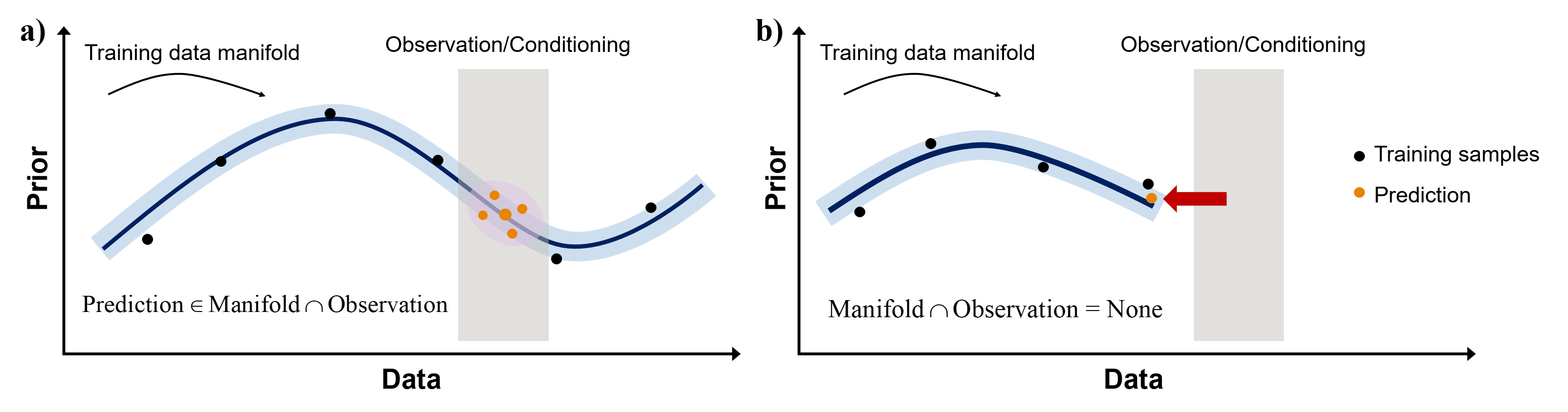}
\caption{Conceptual illustration of the interaction between training data manifold and observation constraints. The horizontal axis represents observation data (e.g., shallow priors, well constraints), while the vertical axis represents the learned prior distribution. Black dots indicate discrete training samples, and the curved surface represents the learned data manifold. The gray vertical band represents observation constraints. (a) In-distribution scenario: the data manifold is sufficiently broad to intersect with observation constraints (light purple region), enabling effective sampling of predictions (orange dots). (b) Out-of-distribution scenario: the narrow data manifold does not intersect with observation constraints; the model generates predictions by finding the nearest point on the manifold (indicated by red arrow).}
\label{fig10}
\end{figure}

\subsection{Data manifold and observation constraints}
To explain the interaction between learned priors and observation data in our depth-progressive framework, we introduce a conceptual interpretation based on the notion of data manifolds. Figure \ref{fig10} illustrates this concept through two scenarios that explain the different behaviors observed in in-distribution and out-of-distribution (OOD) tests. In Figure \ref{fig10}, the horizontal axis represents observation data (e.g., shallow priors, well velocities, or structural constraints), while the vertical axis represents the learned prior distribution. The black dots represent discrete training samples, which constitute a finite and sparse subset of the complete solution space. Through training, the diffusion model not only memorizes these samples but also learns an underlying low-dimensional structure, where we term the training data manifold (depicted by the curved surface). This manifold represents the region of solution space that the model considers geologically plausible. The gray vertical band represents our observation data or conditional constraints.

During depth-progressive VMB with the diffusion model, the generated velocity models should ideally lie at the intersection of the training data manifold and the observation constraints. Since diffusion models are probabilistic, they can sample from this intersection to produce an ensemble of predictions. In Figure \ref{fig10}, the intersection region is indicated by the light purple shading, and the multiple sampled predictions are shown as orange dots.

Figure \ref{fig10}a illustrates the scenario when the training data distribution is sufficiently broad. The corresponding data manifold is wide enough to intersect with the observation constraints, enabling effective velocity model generation. For example, in our in-distribution tests, unconditional generation successfully produces velocity models that match the characteristics of the reference solution when guided solely by the shallow prior. Here, the shallow prior acts as the observation data, guiding the trained model to generate samples that are consistent with both the shallow prior and the learned characteristics from training data with similar features. The model can effectively sample from the intersection region (light purple), producing diverse realizations (orange dots) that all satisfy both the learned geological priors and the observation constraints.

Figure \ref{fig10}b illustrates the scenario when the training data distribution is limited, resulting in a narrow data manifold. In this case, the observation constraints may not intersect with the training data manifold. When faced with this situation, the model seeks the closest point on the manifold to the observation data, as indicated by the red arrow, and generates predictions based on this nearest feature. This behavior explains our observations in the OOD Marmousi test. The shallow prior of the Marmousi section contains a prominent fault feature. Among our training data, the Overthrust model patches contain the most abundant and pronounced fault structures. Consequently, when the model encounters the fault-bearing shallow prior from Marmousi (which lies outside the training distribution), it identifies the Overthrust features as the closest match on the learned manifold and generates models exhibiting Overthrust-like characteristics, even though these differ significantly from the true Marmousi structure.

This manifold perspective provides important insights into both the capabilities and limitations of our approach. When test data fall within or near the training distribution (in-distribution scenarios), the intersection between the learned manifold and observation constraints enables accurate and diverse velocity model generation. However, when test data diverges significantly from the training distribution (OOD scenarios), the model's generations are constrained to the learned manifold, leading to predictions that may not fully capture the true geology. This interpretation underscores the importance of training data diversity for robust generalization and explains why expanding the training distribution would improve performance on diverse geological settings. It also highlights why additional constraints (wells and structures) become increasingly critical in OOD scenarios, as they provide observation data from multiple dimensions that help anchor the model toward the correct solution even when the shallow prior alone is insufficient.

\begin{figure}[htbp]
\centering
\includegraphics[width=0.75\textwidth]{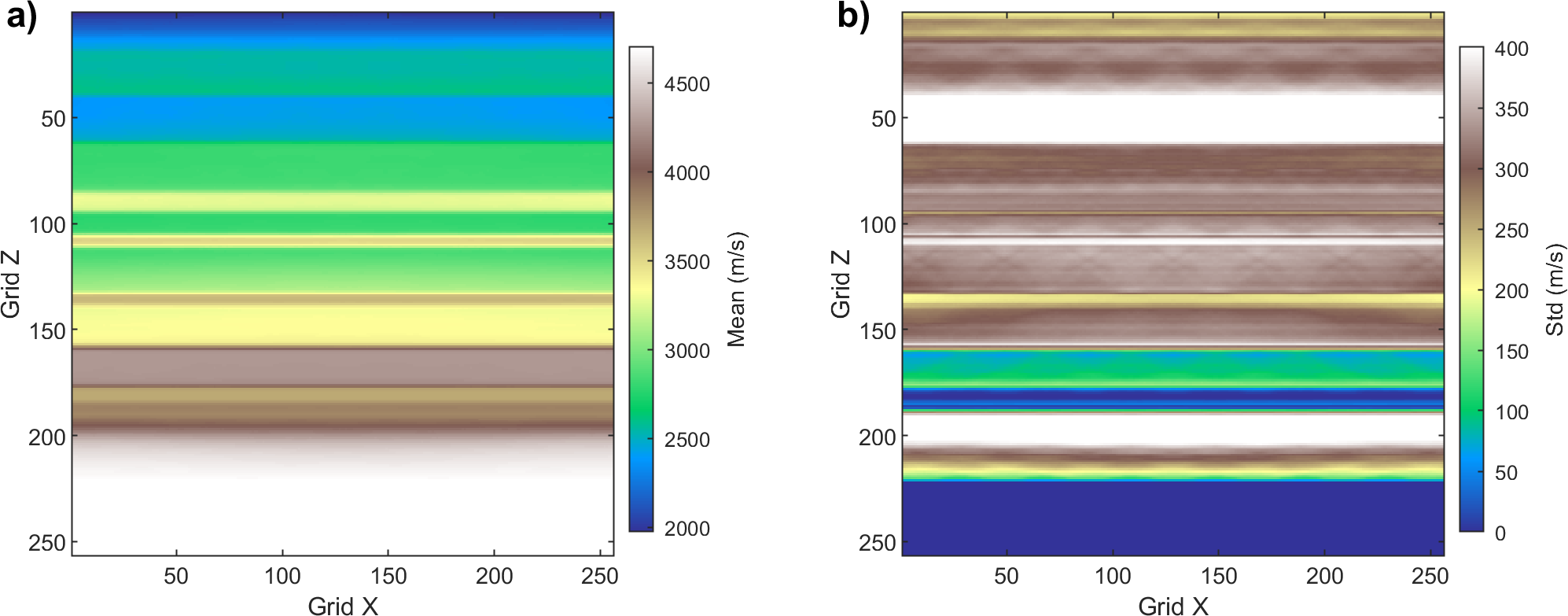}
\caption{Statistical analysis of training data from the Overthrust 3D model. (a) Mean and (b) standard deviation computed across all training patches extracted from the Overthrust 3D volume.}
\label{fig11}
\end{figure}

\subsection{Learned prior distribution vs. Propagated prior distribution}
To concretely illustrate how our method leverages learned statistics and propagates shallow priors, we examine the training data distribution using the Overthrust model as an exemplar. Figure \ref{fig11} presents the mean (panel a) and standard deviation (panel b) computed across all training patches extracted from the Overthrust 3D volume. The mean velocity model indicates a general trend of increasing velocity with depth. The standard deviation map reveals the variability among training patches, where certain depth intervals exhibit high standard deviations. The regions with high variability reflect the diverse structural configurations present in different patches of the Overthrust model. In contrast, relatively homogeneous layers show lower standard deviations, indicating consistent velocity and structure across patches.

These statistical characteristics constitute the prior knowledge learned by the diffusion model during training, influencing the generation process. First, examining the unconditional and well-constrained generation results (Figure \ref{fig6}, columns 1 and 2), we can observe that their mean models (c1, c2) closely match the training data mean (Figure \ref{fig11}a). This indicates that guided by the shallow prior, the diffusion model successfully identifies the Overthrust model's characteristics, and the ensemble of generated samples matches the distribution of Overthrust patches in the training data. This suggests that the shallow prior acted also as a form of classification. In contrast, the structure-constrained and jointly constrained cases (columns 3 and 4) produce mean models (c3, c4) that show deviations from the training data mean. This is expected because these strong constraints heavily narrows the learned prior distribution to match the specific reference model structure, offering an approximate posterior distribution. Second, and more importantly, the learned prior distribution prevents the monotonic accumulation of modeling errors with depth. Conventionally, when building velocity models progressively from shallow to deep, prediction errors would accumulate, leading to increasing uncertainty and errors at greater depths. However, in Figure \ref{fig6}, we can see that both uncertainty (row d) and errors (row e) do not exhibit a general monotonic increasing trend with depth across all four constraint scenarios. For instance, at the model bottom, all four cases show relatively low uncertainty and errors compared to certain shallower regions. This behavior is directly attributable to the learned prior knowledge from training data. From the standard deviation map in Figure \ref{fig11}b, we can observe that the model bottom exhibits low standard deviation, indicating that the structure and velocity distribution at this depth are highly consistent across training patches. This prior knowledge guides the depth-progressive synthesis process, helping the model produce reasonable velocity predictions at specific depths where the training data exhibits low variability, effectively counteracting the error accumulation that would otherwise occur. Of course, this feature is courtesy of using the analytically derived structure as condition. However, it demonstrated that uncertainty is not effected by the shallow to deep progression. This is helpful as we want the uncertainty to be related to the data, and specifically the expected degradation of the seismic image, as a structural constraint, as a function of depth, as we will see in part II.

The role of shallow prior propagation becomes particularly evident when comparing generated uncertainties with training statistics. Despite the high variability in shallow sections of the training data (Figure \ref{fig11}b, $z<50$), all generation scenarios in Figure \ref{fig6} show significantly reduced shallow uncertainty due to the imposed shallow prior constraint. This demonstrates one of the key advantage of our depth-progressive approach: by initiating generation from known shallow velocities and progressively building deeper structures, we effectively override the high uncertainty that would arise from learning shallow velocity distributions alone. The shallow prior acts as a strong anchor point that propagates its influence downward, creating a cascading effect where each depth benefits from the accumulated constraints above.

This interplay between learned distributions and propagated priors creates a robust framework for VMB. The learned statistics provide geological plausibility by encoding typical velocity ranges and variability patterns characteristic of different structural settings, while the shallow prior ensures accurate starting conditions that constrain the entire generation process. When combined with structural and well constraints, this dual mechanism enables our depth-progressive approach to achieve both geological realism and quantitative accuracy in VMB.

\subsection{Impact of shallow prior depth range on velocity model building}
In our depth-progressive synthesis framework, the shallow prior serves as the initial condition that guides the generation of deeper velocity structures. The depth of this shallow prior influences the quality and reliability of the generated velocity models. To systematically investigate this relationship, we conduct a sensitivity analysis using the Overthrust model to examine how varying the shallow prior depth affects unconditional velocity model generation.

\begin{figure}[htbp]
\centering
\includegraphics[width=1\textwidth]{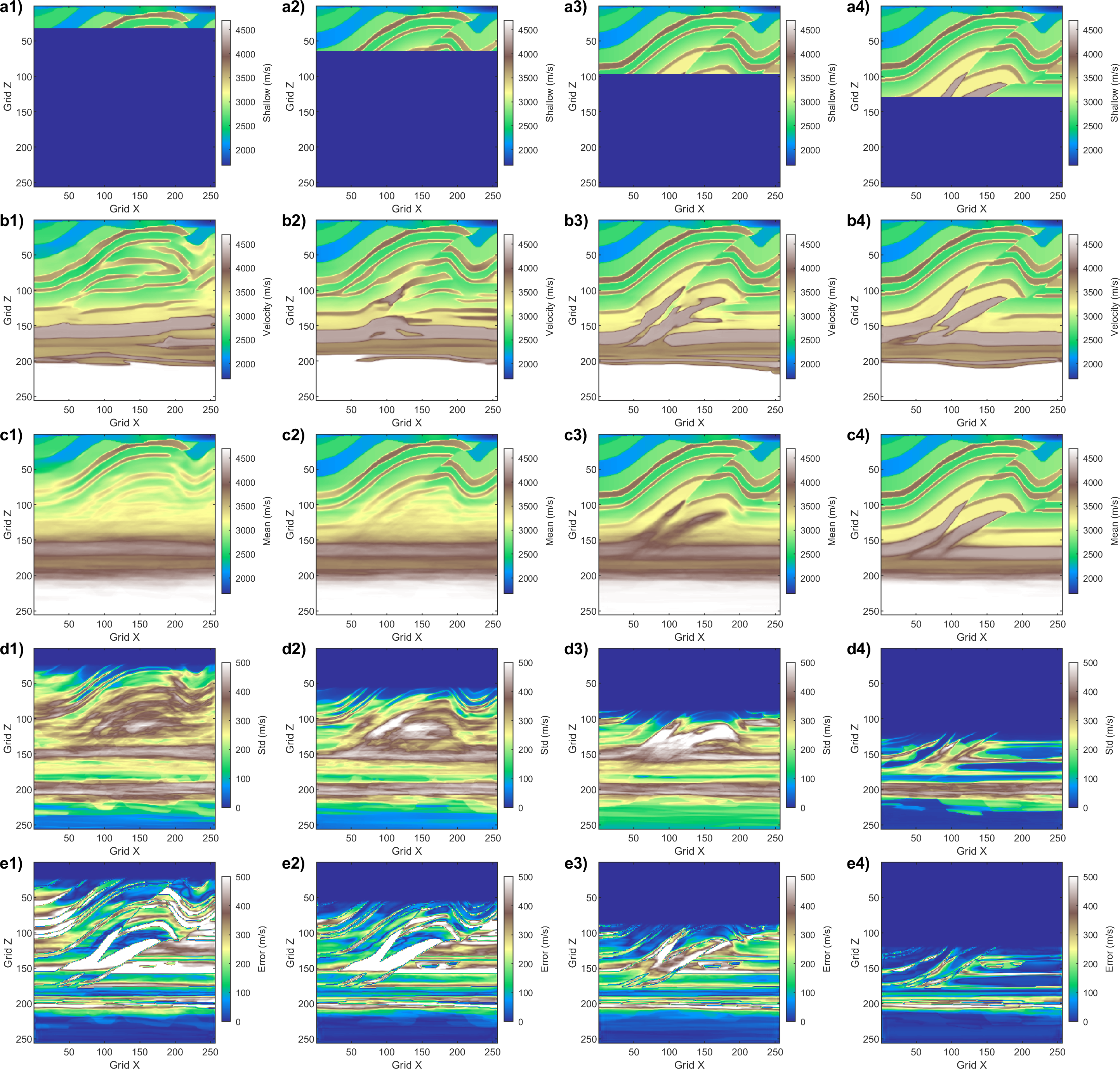}
\caption{Impact of shallow prior depth on unconditional velocity model generation for the Overthrust model. (a1-a4) Shallow priors at depths of 32, 64, 96, and 128 grid points. Rows b-e show single realizations, means, standard deviations, and errors relative to the reference model (Figure \ref{fig6}a1) for each shallow prior depth (columns 1-4).}
\label{fig12}
\end{figure}

Figure \ref{fig12} presents the unconditional velocity model generation results under different shallow prior depth ranges: 32, 64, 96, and 128 grid points, corresponding to columns 1 through 4 respectively. The first row (a1-a4) displays the shallow priors extracted from the reference Overthrust model (Figure \ref{fig6}a1) at different depths. The subsequent rows show the unconditional generation results: single realizations (row b), means of 50 generated realizations (row c), standard deviations (row d), and absolute differences between the mean models and the reference model (row e). The results reveal a clear trend: as the shallow prior depth increases from 32 to 128 grid points, the generated velocity models progressively improve in accuracy. The single realizations and mean models (rows b and c) exhibit increasingly better structural alignment with the reference model. The standard deviation maps (row d) show substantially reduced uncertainty, with high-uncertainty regions progressively confined to deeper sections. The error maps (row e) demonstrate dramatically decreasing velocity differences, with errors becoming minimal when using the deepest 128-point prior. This systematic improvement can be attributed to two factors. First, deeper shallow priors provide richer geological information that more effectively constrains the generation process. Second, the depth-progressive synthesis propagates this information downward layer by layer, and with more initial layers available, the accumulation of prediction errors is reduced.

From a practical perspective, these results suggest that investing in high-quality shallow velocity characterization, whether through shallow seismic surveys, well control in shallow sections, or other near-surface investigation methods, can substantially improve the accuracy of deep VMB. However, there is a trade-off: obtaining accurate velocity information at greater depths typically requires more expensive data acquisition efforts. The results indicate that even moderate increases in shallow prior depth (e.g., from 32 to 64 grid points) can yield significant improvements in modeling accuracy, providing guidance for optimal shallow data acquisition strategies in practical VMB applications.

\section{\textbf{Conclusions}}

We developed a novel depth-progressive diffusion framework for seismic velocity model building (VMB) that integrates the high-fidelity synthesis capabilities of generative diffusion models (GDMs) with an autoregressive, shallow-to-deep propagation scheme. By conditioning each diffusion step on known shallow velocity priors, explicit depth encodings, and auxiliary geophysical constraints, our method overcomes the limitations of one-shot patch-wise generation, like boundary artifacts, lack of depth-dependent confidence, and inability to leverage reliable near-surface information. The Gaussian-blending inference algorithm ensures seamless integration of overlapping patches and yields continuous, geologically plausible velocity volumes. Extensive numerical experiments on both in-distribution tests (SEAM Arid, SEG/EAGE, Overthrust) and a challenging out-of-distribution test (Marmousi II) demonstrate that combining shallow priors with well and structural constraints delivers significant gains in VMB accuracy and substantial uncertainty reduction. Also, the results reveal strong spatial correlation between predicted uncertainties and actual errors, validating the reliability of our uncertainty quantification approach for practical VMB applications.

However, as a proof of concept, this part employs structural constraints derived from vertical reflectivity computed directly from reference velocity models, representing an idealized scenario that demonstrates the potential of structural conditioning in our framework. While these ideal constraints effectively validate the methodological framework and establish the fundamental capabilities of depth-progressive diffusion for VMB, they are typically unavailable in real-world exploration settings. To bridge this gap and demonstrate the practical viability of our approach, the companion paper (Part II) extends the framework to incorporate realistic structural constraints derived from migrated images and validates the method on field seismic data, establishing the operational feasibility and industrial relevance of depth-progressive diffusion framework for seismic VMB.

\section{Acknowledgments}
This publication is based on work supported by the King Abdullah University of Science and Technology (KAUST). The authors thank the DeepWave sponsors for their support. This work utilized the resources of the Supercomputing Laboratory at King Abdullah University of Science and Technology (KAUST) in Thuwal, Saudi Arabia.
\section{Code and data availability}
All codes, datasets, and pre-trained models associated with this work are publicly available to ensure full reproducibility of the results reported in the manuscript.

\subsection{Code availability}

The source code is openly available on GitHub at \url{https://github.com/DeepWave-KAUST/DiffVMB-pub}. The repository includes the complete implementation of the depth-progressive diffusion framework, organized into two parts corresponding to the two companion manuscripts. For Part~I, the \texttt{diffvmb\_part1/} directory contains the core Python library (\texttt{code/}), the training script (\texttt{train.py}), and the sampling/inference script (\texttt{sample.py}). The framework is built upon the IDDPM architecture \citep{ho2020denoising, nichol2021improved} and extended with custom multi-condition inputs, including shallow velocity context, depth positional encoding, well-log constraints, and reflectivity-based structural constraints. Installation instructions and dependencies are provided in the repository README.

\subsection{Dataset availability}

The training and test datasets are publicly available on Zenodo \citep[DOI:][]{zenodo2025diffvmb} as part of the archive \texttt{dataset.zip}. After extraction, the Part~I data are located under \texttt{dataset/part1/}, which contains two sub-directories:
\begin{itemize}
    \item \texttt{train/}: training samples in NPZ format, each containing two arrays, namely \texttt{vp} (P-wave velocity, shape $n_z \times n_x$) and \texttt{ref} (reflectivity, same shape), which represents 2-D sections extracted from industrial velocity models;
    \item \texttt{test/}: four benchmark velocity models in MAT format used for evaluation, comprising three in-distribution models (SEAM Arid, SEG/EAGE, and Overthrust) and one out-of-distribution model (Marmousi) to assess generalization capability.
\end{itemize}

\subsection{Pre-trained model}

The pre-trained model weights for Part~I are available on the same Zenodo record \citep[DOI:][]{zenodo2025diffvmb} as part of the archive \texttt{trained\_model.zip}. After extraction, the weights are stored in \texttt{trained\_model/model\_part1.pt}. The model was trained on a single NVIDIA A100 GPU. To reproduce the numerical results reported in this manuscript, users may download the pre-trained weights and the test dataset, place them in the directories specified in the repository README, and execute:
\begin{verbatim}
cd diffvmb_part1
python sample.py
\end{verbatim}

\bibliographystyle{unsrtnat}
\bibliography{references}

\end{document}